\documentclass[aps,prb,showpacs,groupedaddress,superscriptaddress,twocolumn]{revtex4}
\usepackage{amsfonts}
\usepackage{graphicx}
\usepackage{latexsym}
\usepackage{makeidx} 
\usepackage{amsmath} 
\usepackage{amssymb}
\usepackage{graphicx}
\usepackage{color}
\begin{document}

\title{Nonequilibrium theory of the conversion-efficiency limit of solar cells including thermalization and extraction of carriers}
\author{Kenji Kamide}
\email{kenji.kamide@aist.go.jp}
\affiliation{%
Renewable Energy Research Center, National Institute of Advanced Industrial Science and Technology (AIST), Koriyama, Fukushima, 963-0298, Japan}
\author{Toshimitsu Mochizuki}
\affiliation{%
Renewable Energy Research Center, National Institute of Advanced Industrial Science and Technology (AIST), Koriyama, Fukushima, 963-0298, Japan}
\author{Hidefumi Akiyama}%
\affiliation{%
The Institute for Solid State Physics (ISSP), The University of Tokyo, Kashiwa, Chiba 277-8581, Japan}
\affiliation{%
OPERANDO-OIL, Kashiwa, Chiba 277-8581, Japan}
\author{Hidetaka Takato}
\affiliation{%
Renewable Energy Research Center, National Institute of Advanced Industrial Science and Technology (AIST), Koriyama, Fukushima, 963-0298, Japan}
\date{\today}
\begin{abstract}
The ideal solar cell conversion efficiency limit known as the Shockley-Queisser (SQ) limit, which is based on a detailed balance between absorption and radiation, has long been a target for solar cell researchers. 
While the theory for this limit uses several assumptions, the requirements in real devices have not been discussed fully.
Given the current situation in which research-level cell efficiencies are approaching the SQ limit, a quantitative argument with regard to these requirements is worthwhile in terms of understanding of the remaining loss mechanisms in current devices and the device characteristics of solar cells that are operating outside the detailed balance conditions. 
Here we examine two basic assumptions: (1) that the photo-generated carriers lose their kinetic energy via phonon emission in a moment (fast thermalization), and (2) that the photo-generated carriers are extracted into carrier reservoirs in a moment (fast extraction). Using a model that accounts for the carrier relaxation and extraction dynamics, we reformulate the nonequilibrium theory for solar cells in a manner that covers both the equilibrium and nonequilibrium regimes. Using a simple planar solar cell as an example, we address the parameter regime in terms of the carrier extraction time and then consider where the conventional SQ theory applies and what could happen outside the applicable range.
\end{abstract} 
\maketitle 
\noindent 

\section{\label{SecI}Introduction}
Shockley and Queisser (SQ) determined a theoretical estimate for the upper limit of the conversion efficiency in an ideal solar cell~\cite{SQ}.
The original SQ theory takes radiative recombination into account as a main cause of the current loss in solar cells in a simple manner.
The energy distributions of the carriers, denoted by $n^e_{E_e}$ for electrons and $n^h_{E_h}$ for holes, must be known to evaluate the radiative recombination rate because it is proportional to the sum of their product. 
SQ theory assumes that the carriers in the absorber are in thermal and chemical equilibrium with both the lattice phonons and the carriers in the electrodes at an ambient temperature $T_c$ and with chemical potentials of $\mu_c$ for the conduction electrons and $\mu_v$ for the valence electrons.
The resulting current-voltage relationship given by $I=I(V)=I_{\rm sun}-I_{\rm rad}^0 e^{|{\rm e}|V/k_{\rm B}T_c}$, where the voltage between the electrodes is equal to the Fermi level separation within the absorber, $|{\rm e}|V=\mu_c-\mu_v$, ultimately determines the conversion efficiency during maximum power operation, where $I_{\rm sun}$ is the photo-generated current produced by absorption of sunlight.
The detailed balance, which is the essential aspect of the SQ assumptions, has been used routinely in later analyses that further incorporated various additional factors (including Auger recombination, light trapping, photon recycling, and Coulomb interactions ~\cite{Richter}).  

The requirements for the assumptions used in the SQ theory to be justified are commonly described as follows: 
\begin{enumerate}
\renewcommand{\labelenumi}{\arabic{enumi})}
 \item{\label{req1}} the photo-generated carriers lose their kinetic energy via phonon emission and rapidly establish their thermal equilibrium distribution in a moment (which is called fast thermalization);
 \item{\label{req2}} the carriers are extracted rapidly into carrier reservoirs immediately after they are produced (which is called fast extraction).
\end{enumerate}
The latter assumption \ref{req2}) is actually given explicitly in the original paper ~\cite{SQ}. 
However, the above requirements are not sufficiently clear and thus some quantitative issues remain. 
While the two time scales, i.e., the carrier thermalization time, $\tau_{\rm ph}$, and the carrier extraction time, $\tau_{\rm out}$, are assumed to be short, the following questions are not addressed:
first, how short should these times be, i.e., which timescales from other processes should be compared with these times, and 
second, how do $\tau_{\rm ph}$ and $\tau_{\rm out}$ compare?  
The latter question relates directly to the concept of hot carrier solar cells operating out of equilibrium~\cite{Nozik, Wurfel2, Takeda1, Takeda2, Suchet}, where fast carrier extraction before the thermalization is complete can reduce the thermalization losses and ensure that device performance is not limited by a detailed balance.

Record efficiencies of recent cell research are gradually approaching the SQ limits in nonconcentrator-type single-junction solar cells, e.g., Kaneka's Si-based cell with 26.7 percent efficiency and Alta Devices' thin-film GaAs-based cell with 28.8 percent efficiency~\cite{EffTab50}.
It is therefore important to have a more precise understanding of the situation in which detailed balance theory provides a reliable estimate of the attainable upper efficiency limit.
Quantitative estimation of the parameters to which the SQ theory applies will help to clarify the remaining energy losses and push current device performance towards the SQ limit.
Additionally, a more precise understanding of the energy conversion mechanisms from the detailed balance will lead to new strategies for future improvements that are intended to go beyond the SQ limit.

The nonequilibrium dynamics of many particle systems can be described in general terms using nonequilibrium Green's functions (NEGFs)~\cite{Martin-Schwinger, Kadanof-Baym, Keldysh}.
These functions were initially applied to study electron transport in solids and in mesoscopic devices~\cite{Rammer, Datta}, and later in semiconductor light-emitting devices (e.g. light-emitting diodes or LEDs~\cite{Steiger}, semiconductor lasers~\cite{Henneberger}, quantum cascade lasers~\cite{Lee}, and polariton condensates~\cite{Szymanska, Yamaguchi}). 
More recently, the NEGF formalism was also used to study solar cells with nanostructured absorbers~\cite{Aeberhard1, Aeberhard2, Cavassilas}, where the device characteristics are likely to be affected strongly by the quantum transport of the carriers.
NEGFs were also used to study the conditions required to validate use of luminescence-based characterization of solar cells~\cite{Aeberhard3}, which is justifiable in terms of photovoltaic reciprocity under the detailed balance principle~\cite{Rau, Kirchartz}. 
Despite the sound theoretical basis that is available, the device characteristics have not been explored for a sufficiently wide range of parameters via the NEGF approach, particularly for solar cells. This seems to be related to the complexity of the theory and high computational costs.  
Similar issues were found with an {\it ab initio} approach~\cite{Bernardi}

In this work, we present a nonequilibrium theory that does not assume any form for the distribution functions used for the carriers in the absorber, in a manner similar to the NEGF formulation.
The carrier distribution functions in the absorber are determined using a set of rate equations that is derived from second-order perturbation theory based on the coupling between the absorber carriers and three baths (the phonon, electron, and hole reservoirs). Spectral broadening of the microscopic states of the carriers is also included in the relevant cases. 
As a result, the theory describes solar cell operation for a wide range of parameters, including the situations where the photo-generated carriers are either in or out of thermal equilibrium.  

\begin{figure}[!t]
\centering
\includegraphics[width=0.45\textwidth]{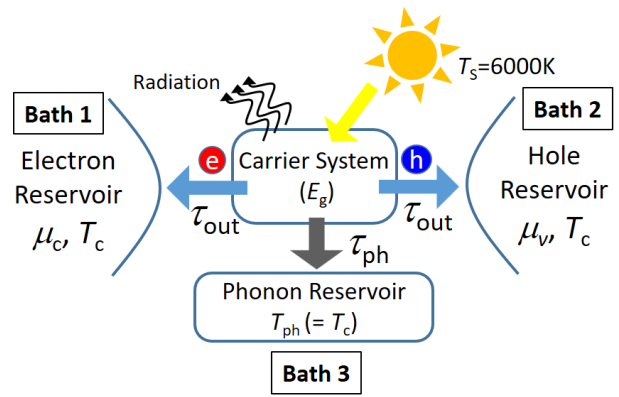}
\caption{Nonequilibrium model of solar cell: electron and hole carriers in a semiconductor absorber (energy gap, $E_g$) interact with the carriers in the carrier reservoirs (Bath 1 and Bath 2) and the lattice phonons in the absorber (Bath 3). 
Population distributions of the carriers and phonons in the three reservoirs are given using the thermal equilibrium distribution function (Fermi-Dirac distribution with chemical potentials of $\mu_c$ and $\mu_v$ for conduction and valence electrons, respectively, at temperature $T_c (=300 K)$, and Bose-Einstein distribution for phonons at lattice temperature $T_{\rm ph}(=T_c)$). 
Electron-hole pairs are generated in the absorber by solar ilumination (a blackbody spectrum at temperature $T_S (=6000 K)$ is assumed here), and the output current loss is due to radiative recombination in the absorber.}  
\label{fig1}   
\end{figure}

This paper is organized as follows.
In section \ref{SecII}, we formulate a nonequilibrium theory for solar cells based on the model shown in Fig.~\ref{fig1}, and derive a set of rate equations for the microscopic carrier distribution function in the absorber.
The microscopic carrier distribution function is then determined as a steady-state solution to the rate equation.
At the end of this section, general expressions are given for the total output current and the total output power to enable simulation of the solar cell device performance.  
In section \ref{SecIII}, the basic properties required by the solution to the set of rate equations are presented before the numerical analysis begins.
These properties are useful when verifying the accuracy of the simulation. 
Classification of the parameter regime, specifically in terms of the carrier extraction time $\tau_{\rm out}$, is also presented in Sec.~\ref{SecIII}.
Before the set of equations is solved, the equations themselves can be used to indicate the parameter regime where the assumptions of the SQ theory fail.
In section \ref{SecIV}, a device performance simulation based on our formulation is presented for a simple planar single-junction solar cell.
The numerical simulations show what physically happens in the photovoltaic energy conversion processes in each of the regimes that were classified in Sec.~\ref{SecIII}.
In section~\ref{SecV}, we summarize these findings and discuss future issues and future applications of the nonequilibrium theory.

Finally, in this section, we list definitions for the symbols used in this paper. 
Parameters for bulk semiconductors can be found in the standard textbook~\cite{Cardona}.
\begin{itemize} 
\item $c=$ speed of light $=3 \times 10^8 $ m/s
\item $\hbar=$Planck constant/2$\pi =1.0545718 \times 10^{-34} $ J s
\item $R_S=$ Sun's radius $=0.696 \times 10^6$ km
\item $L_{ES}=$ average distance from the Earth to the Sun $=1.496 \times 10^8$ km
\item $\mathcal{CR}\left(\le(L_{ES}/R_S)^2=46200 \right) =$ concentration ratio
\item $w=$ absorber thickness in a planar solar cell
\item $\mathcal{A}=$ absorber area in a planar solar cell
\item $\mathcal{V}=\mathcal{A} w =$ absorber volume in a planar solar cell
\item $T_S=$ surface temperature of the Sun $=6000$ K
\item $T_c=$ ambient temperature = room temperature $=300$ K
\item $T_{\rm ph}(=T_c)=$ absorber lattice temperature 
\item $k_B=$ Boltzmann constant $=8.6 \times 10^{-5}$ eV/K
\item $\beta_S =1/(k_B T_S)$, $\beta_c =1/(k_B T_c)$, $\beta_{\rm ph} =1/(k_B T_{\rm ph})$
\item $m^\ast_{e(h)}=$ effective mass of electrons (holes) in the absorber (Si: $m^\ast_e/m_e=(\nu_{\rm valley}^2m_{\perp}^2 m_\parallel)^{1/3}/m_e = 1.08$, $m^\ast_h/m_e=(m_{hh}^{3/2}+m_{lh}^{3/2})^{2/3}/m_e=0.55$ with $\nu_{\rm valley}=6$, $m_{\perp}/m_e=0.19$, $m_{\parallel}/m_e=0.98$, $m_{hh}/m_e=0.49$, $m_{lh}/m_e=0.16$)
\item $m_e=$ bare electron mass = 9.1 $\times 10^{-31}$ kg
\item $\mathcal{D}_{e(h)}(E_{e(h)})=d_{e(h)} \sqrt{E_{e(h)}}=$ density of states per unit volume for electrons (holes) in the absorber with kinetic energy $E_e$ ($E_h$) where $d_{e(h)}(=\frac{(2 m^\ast_{e(h)})^{3/2}}{2 \pi^2 \hbar ^3} )$
\item $E_g=$ absorber bandgap (Si: 1.12 eV, GaAs: 1.42 eV)
\item $\tau_{\rm out}=$ carrier extraction time
\item $\mu_c=$ Fermi level of electrons in electron reservoir (Bath 1) $=E_g/2+\beta_c^{-1}/2 \ln \frac{d_h}{d_e}+|e|V/2$ (charge neutrality condition)
\item $\mu_v=$ Fermi level of electrons in hole reservoir (Bath 2) $=E_g/2+\beta_c^{-1}/2 \ln \frac{d_h}{d_e}-|e|V/2$ (charge neutrality condition)
\item $g_q^{c(v)}=$ electron-phonon coupling constant for conduction (valence) band carriers ($=a_{{\rm def}, c} \sqrt{\frac{\hbar q}{2V v_A \rho_A }}$ for longitudinal-acoustic (LA) phonons with $q$=phonon wave number)
\item $a_{{\rm def}, c}=$ deformation potential for electrons in the bottom conduction band in the absorber (Si: $\sim 10$ eV)
\item $a_{{\rm def}, v}=$ deformation potential for electrons in the top valence band in the absorber (Si: $\sim a_{{\rm def}, c}/10 \sim 1$ eV)
\item $v_A=$ LA phonon velocity (Si: $\sim 10^4$ m/s)
\item $\rho_A=$ absorber mass density (Si: 2.3 g/cm$^3$)
\item $f^{F(B)}_{\mu,\beta}(E)=$ Fermi-Dirac (Bose-Einstein) distribution at chemical potential $\mu$ and inverse temperature $\beta$.
\end{itemize} 

\section{\label{SecII}Nonequilibrium theory (Formulation)}
\begin{figure}[!t]
\centering
\includegraphics[width=0.45\textwidth]{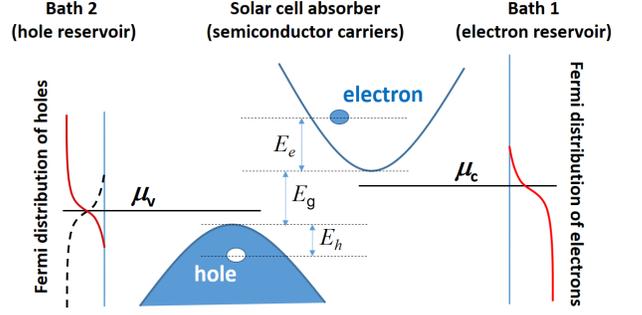}
\caption{Energy diagrams for carriers in the absorber and Fermi distribution functions in the carrier reservoirs (Baths 1 and 2).}  
\label{fig2}   
\end{figure}
\begin{figure}[!t]
\centering
\includegraphics[width=0.49\textwidth]{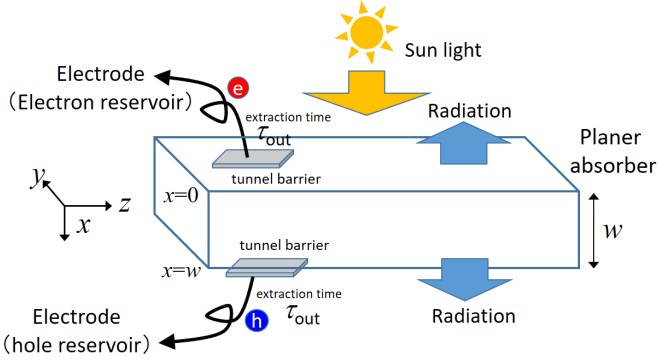}
\caption{Simple planar solar cell model: a planar absorber of thickness $w$ is illuminated by sunlight from the normal direction and photo-generated carriers are extracted 
to the two electrodes (Baths 1 and 2) through tunneling contact. The photocarrier losses are due to radiative recombination.}  
\label{fig3}  
\end{figure}

In this section, we formulate the nonequilibrium theory for solar cells.
As shown below, a set of rate equations for the microscopic distribution functions for the electrons ($e$) and holes ($h$) in the absorber, ($\{ n^e_{E_e}, n^h_{E_h} \}$), are given in the following form:
\begin{eqnarray}
\frac{d}{dt}n^e_{E_e}&=&J^{e,{\rm sun}}_{E_e}-J^{e,{\rm rad}}_{E_e}-J^{e, {\rm out}}_{E_e}+\left.\frac{d}{dt}n^e_{E_e}\right|_{\rm phonon}, \label{eq:e-balance}  \quad \\
\frac{d}{dt}n^h_{E_h}&=&J^{h,{\rm sun}}_{E_h}-J^{h,{\rm rad}}_{E_h}-J^{h, {\rm out}}_{E_h}+\left.\frac{d}{dt}n^h_{E_h}\right|_{\rm phonon}. \label{eq:h-balance}  \quad
\end{eqnarray}
Here, $E_e$ and $E_h$ are the carrier kinetic energies measured from the bottom of the bands (Fig.~\ref{fig2}). 
The first term, $J^{e(h),{\rm sun}}_{E_{e(h)}}$, on the right-hand side of the rate equations represents the carrier generation rate due to sunlight absorption.
The second term, $J^{e,{\rm rad}}_{E_e(h)}$, represents the carrier loss rate due to radiative carrier recombination.
The third term, $J^{e(h),{\rm out}}_{E_{e(h)}}$, represents the rate of carrier extraction to the electrodes.
The last term, $\left.\frac{d}{dt}n^{e(h)}_{E_{e(h)}}\right|_{\rm phonon}$, represents the rate of electron scattering to other microscopic states within the same band due to phonon emission or absorption. 
For the solar cell characteristics simulation, this equation will be solved under the steady-state condition:
\begin{eqnarray}
\frac{d}{dt}n^e_{E_e}=\frac{d}{dt}n^h_{E_h}=0.
\end{eqnarray}
In the following subsections, we will derive explicit expressions for, $J^{e(h),{\rm sun}}_{E_{e(h)}}$, $J^{e(h),{\rm rad}}_{E_{e(h)}}$, $J^{e(h),{\rm out}}_{E_{e(h)}}$, and $\left.\frac{d}{dt}n^{e(h)}_{E_{e(h)}}\right|_{\rm phonon}$ via microscopic modeling of the carriers in the simple planar solar cell (thickness $w$, surface area $\mathcal{A}$, volume $\mathcal{V}=\mathcal{A}w$) shown in Fig.~\ref{fig3}.

The assumptions that were made in the original SQ model~\cite{SQ} are also used here to simplify the discussion but will not alter the main conclusion of this paper.
For example, we consider the absorber thickness $w$ to be larger than the absorption length but less than the minority carrier diffusion length. This allows us to consider perfect absorption of sunlight above the absorption edge $(E>E_g)$ and a homogeneous carrier distribution in the absorber. Perfect anti-reflection behavior at the front surface and perfect passivation with zero surface recombination are also assumed here.

An additional simplification is made in this work to the band structure of the carriers in the absorber. 
An effective two-band model is used to describe the microscopic carrier states under an effective mass approximation (with infinite bandwidths), in which the effective masses for the electrons ($m_e^\ast$) and holes ($m_h^\ast$) were selected to reproduce the densities of states near the band extrema. 
Therefore, the effects of the band anisotropy, the valley degree of freedom within the degenerate bands (particularly in Si), and the contributions from other bands located away from the extrema were not taken into account correctly. 
In this sense, our analysis is far from but is not intended to be quantitatively accurate in simulations for specific systems, but is rather intended to produce a general picture of the main issue, i.e., nonequilibrium aspects of solar cells.

\subsection{\label{Jsun} Generation rate due to sunlight absorption: $J^{e(h),{\rm sun}}_{E_{e(h)}}$}
\begin{figure}[!t]
\centering
\includegraphics[width=0.49\textwidth]{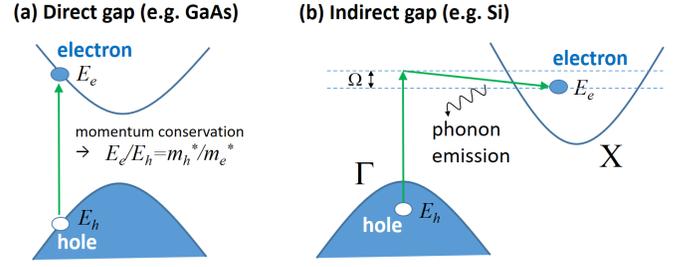}
\caption{Absorption processes in (a) direct and (b) indirect gap semiconductors.}  
\label{fig4}  
\end{figure}

Under the assumption of perfect absorption (where $w \gg$ absorption length), the number of photons absorbed into the absorber per unit time through surface area $\mathcal{A}$ is given by
\begin{eqnarray}
\mathcal{A} \times \int_{E_g}^{\infty}  j^{\rm sun}(E) {\rm d}E \label{eq:e-gen-sun-total}.
\end{eqnarray}
The solar spectrum for the photon number current (per unit area, per unit time, and per unit energy), $j^{\rm sun}(E)$, is simply approximated using blackbody radiation at $T_S=6000 K$ under the AM0 condition~\cite{Wurfel text},
\begin{eqnarray}
j^{\rm sun}(E)= \mathcal{CR}  \times \frac{c}{4}\left( \frac{R_S}{L_{ES}}\right)^2   \mathcal{D}^0_{\gamma}(E) \times f^B_{0,\beta_S} (E) , \label{eq:AM0blackbody}
\end{eqnarray}
where $\mathcal{CR} $ is the concentration ratio, $\mathcal{D}^0_{\gamma}(E)=\frac{1}{3\pi^2} ( \hbar c)^{-3}\times 3 E^2$ is the photonic density of states in a vacuum, and $f^B_{0,\beta_S} (E)=(\exp(\beta_S E)-1)^{-1}$ is the Bose-Einstein distribution function at energy $E$ with inverse temperature $\beta_S=1/(k_B T_S)$. 
If necessary, for practical device simulations, the solar spectrum may be replaced appropriately, e.g., using the AM1.5 spectrum normalized at a total power of 1 kWm$^{-2}$, which is not the case here (the 6000 K blackbody spectrum in Eq.~(\ref{eq:e-gen-sun-total}) and Eq.~(\ref{eq:AM0blackbody}) approximates the AM0 spectrum at a total power of 1.6 kW/m$^2$ at 1 sun, with $\mathcal{CR}=1$).
In the rate equations in Eq.~(\ref{eq:e-balance}) and Eq.~(\ref{eq:h-balance}), the generation rates of the microscopic carrier distribution function $J^{e(h),{\rm sun}}_{E_{e(h)}}$ should be expressed using the solar spectrum, $j^{\rm sun}(E)$.  
The expression is dependent on whether the absorber is made from direct or indirect gap semiconductors (Fig.~\ref{fig4}).

{\it For direct gap semiconductor absorbers} (Fig.~\ref{fig4} (a)) ---
by considering momentum and energy conservations, we can equate the number of carriers that are generated in energy ranges of $E_e<E'<E_e+{\rm d} E_e$ for electrons and $E_h<E'<E_h+{\rm d} E_h$ for holes with the number of photons absorbed in the energy range $E<E'<E+{\rm d} E$ per unit time in the absorber as follows:
\begin{eqnarray}
\mathcal{A} j^{\rm sun}(E) {\rm d}E &=&\mathcal{D}_e(E_e) \mathcal{V}  J^{e, {\rm sun}}_{E_e}  {\rm d}E_e \nonumber \\
&=&\mathcal{D}_h(E_h) \mathcal{V}   J^{h, {\rm sun}}_{E_h}  {\rm d}E_h,
\end{eqnarray}
where the energy conservation law gives $E=E_g+E_e+E_h$, and momentum conservation under the effective mass approximation gives $E_e=\frac{m_h^\ast}{m_e^\ast}E_h$.
Here $\mathcal{D}_{e}(=d_e\sqrt{E_e})$ and $\mathcal{D}_{h}(=d_h\sqrt{E_h})$ are the densities of states of electrons and holes per unit volume, respectively.
The equation relates $J^{e(h), {\rm sun}}_{E_e(h)}$ and $j^{\rm sun}(E)$ directly, as follows:
\begin{eqnarray}
&&J^{e, {\rm sun}}_{E_e} =\frac{ j^{\rm sun}\left(E=E_g+(1+\frac{m_e^\ast}{m_h^\ast})E_e \right) }{w\mathcal{D}_e(E_e)}  \left(1+\frac{m_e^\ast}{m_h^\ast}\right), \quad \label{eq:Je-sun-direct} \\
&&J^{h, {\rm sun}}_{E_h} =\frac{j^{\rm sun}\left(E=E_g+(1+\frac{m_h^\ast}{m_e^\ast})E_h \right) }{w\mathcal{D}_h(E_h)}  \left(1+\frac{m_h^\ast}{m_e^\ast}\right), \quad \label{eq:Jh-sun-direct}
\end{eqnarray}
where we assume that all microscopic states of carriers with the same energy are generated with equal probability, independent of their momentum directions. We therefore assume that the carrier distribution function is solely dependent on the kinetic energy of carriers and independent of the momentum direction. This assumption is used throughout the paper.

{\it For indirect gap semiconductor absorbers} (Fig.~\ref{fig4} (b)) --- the absorption process accompanies photon emission or absorption. 
The energies of the electron-hole pairs deviate from the photon energy by the energy of one phonon ($(\Omega_{LA},\Omega_{TA},\Omega_{TO})=(50.9, 57.4, 18.6)$ meV for Si). 
We simply neglect the energy shift here because the phonon energy is much smaller than the spectral bandwidth, $\sim k_B T_S$, of the incoming sunlight.
However, because the indirect transition accompanies a shift in the carrier momentum corresponding to the momentum carried by the phonons, momentum conservation among the photon and electron-hole pairs is not required in the absorption process.
As a result, electron-hole pairs with arbitrary combinations of the energies $E_e$ and $E_h$ can be created by absorption of one photon with energy $E$, as long as $E=E_g+E_e+E_h$ is satisfied (where the phonon energy shift is neglected). 
The situation in indirect gap semiconductors means that the expressions for $J^{e(h), {\rm sun}}_{E_e(h)}$ from Eq.~(\ref{eq:Je-sun-direct}) and Eq.~(\ref{eq:Jh-sun-direct}) for direct gap semiconductors must be altered.
Assuming that all electron-hole pairs with $E_e (<E-E_g)$ and $E_h(=E-E_g-E_e)$ are created by absorption of one photon with energy $E$ with equal probability, the probability $p^e_{E_e}(E){\rm d}E_e$ of finding electrons in a small energy window $E_e<E'<E_e+{\rm d}E_e$ immediately after absorption is
\begin{eqnarray}
p^e_{E_e}(E){\rm d}E_e=\frac{\mathcal{D}_e(E_e)\mathcal{D}_h(\Delta E-E_e) {\rm d}E_e }{\int_0^{\Delta E} \mathcal{D}_e(E')\mathcal{D}_h(\Delta E-E') {\rm d}E'}, \label{eq:pe}
\end{eqnarray}
where $\Delta E \equiv E-E_g$. Because the number of photons absorbed in the absorber per unit time and per unit energy is $\mathcal{A} j^{\rm sun}(E)$ for $E>E_g$, 
we find the following expression for the generation rate of electrons per microscopic state for indirect gap semiconductor absorbers: 
\begin{eqnarray}
J^{e, {\rm sun}}_{E_e} &=&\frac{\int_{E_g}^{\infty} \mathcal{A} j^{\rm sun}(E)   \times  p^e_{E_e}(E) {\rm d}E_e {\rm d}E  }{\mathcal{V} \mathcal{D}_e(E_e) {\rm d}E_e } \nonumber \\
&=&\int_{E_g+E_e}^{\infty} \frac{ j^{\rm sun}(E)\times \mathcal{D}_h(\Delta E-E_e)/w}{\int_0^{\Delta E} \mathcal{D}_e(E')  \mathcal{D}_h(\Delta E-E'){\rm d}E'}{\rm d}E \nonumber \\
&=&\int_{0}^{\infty} \frac{ j^{\rm sun}(E_g+E_e+E_h)\times \sqrt{E_h}}{\pi w d_e  (E_e+E_h)^2/8 }{\rm d}E_h.   \label{eq:Je-sun2}
\end{eqnarray}
The hole generation rate is given in a similar manner as
\begin{eqnarray}
J^{h, {\rm sun}}_{E_h}&=&\int_{0}^{\infty} \frac{ j^{\rm sun}(E_g+E_e+E_h)\times \sqrt{E_e}}{\pi w d_h  (E_e+E_h)^2/8 }{\rm d}E_e. \label{eq:Jh-sun2}
\end{eqnarray}

\subsection{\label{Jrad} Recombination loss rate: $J^{e(h),{\rm rad}}_{E_{e(h)}}$}
\begin{figure}[!t]
\centering
\includegraphics[width=0.49\textwidth]{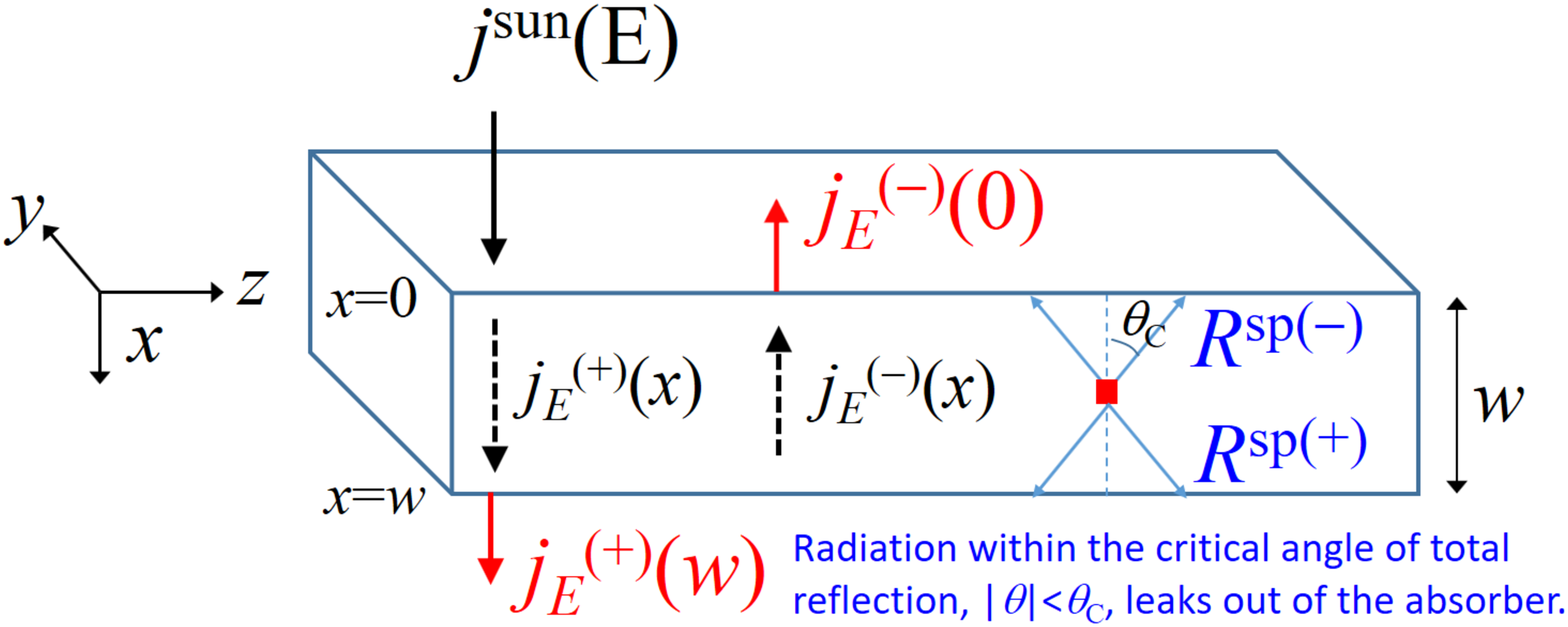}
\caption{Schematic for calculation of recombination loss rate, $J^{e(h),{\rm rad}}_{E_{e(h)}}$}  
\label{fig5} 
\end{figure}

The derivation of the expression for $J^{e(h),{\rm rad}}_{E_{e(h)}}$ presented in this subsection largely follows the derivations in the literature~\cite{Wurfel1, Wurfel text}.
Because the absorber thickness $w$ considered here is much smaller than the minority carrier diffusion length, we can safely assume a homogeneous carrier distribution inside the absorber.
For a given set of carrier distribution functions, $\{ n^e_{E_e}, n^h_{E_h} \}$, the recombination radiation rate of photons $R^{sp}(E)$ at photon energy $E$ from the arbitrary position of a small volume inside the absorber into the whole solid angle ($4\pi$), per unit volume, per unit energy, and per unit time, is   
\begin{eqnarray}
R^{sp}(E)&=&(\frac{c}{n})|\mathcal M|^2\mathcal{D}_{\gamma}^{\rm cell}(E) \\
 &\times& \int_0^{\Delta E} \mathcal{D}_e(E')\mathcal{D}_h(\Delta E-E') n^e_{E'} n^h_{\Delta E-E'} {\rm d}E', \nonumber 
\end{eqnarray}
for an indirect gap semiconductor absorber. Here, $\Delta E \equiv E-E_g$, $\frac{c}{n}$ is the speed of light inside the absorber with refractive index $n$, $\mathcal{M}$ is proportional to the phonon-mediated transition matrix element that is approximated using the value at the absorption edge, and $\mathcal{D}_{\gamma}^{\rm cell}(E)=\frac{1}{3\pi^2} ( \hbar c/n)^{-3}\times 3 E^2$ is the photonic density of states in the absorber.
It is important to note that only part of the radiation, i.e., the radiation into the limited solid angle within the critical angle of total reflection, can escape from the absorber, as shown in Fig.~\ref{fig5}, and this results in the photovoltaic current loss.
The radiation rate $R^{sp(\pm)}(E)$ into the escape cones in the $\pm x$-direction, per unit volume, per unit energy, and per unit time, is then,
\begin{eqnarray}
&&R^{sp(\pm)}(E)= \langle |({\vec c}/n)\cdot {\vec e}_{\pm x} | \rangle_{\theta<\theta_c} \times |\mathcal M|^2\mathcal{D}_{\gamma}^{\rm cell}(E)   \nonumber  \\
&& \times \int_0^{\Delta E} \mathcal{D}_e(E')\mathcal{D}_h(\Delta E-E') n^e_{E'} n^h_{\Delta E-E'} {\rm d}E', \quad
\end{eqnarray}
where $ \langle |\frac{{\vec c}}{n}\cdot {\vec e}_{\pm x} | \rangle_{\theta<\theta_c} $ represents the velocity of light in the $\pm x$-directions when averaged inside the escape cones.
The geometric average yields an expression for the rates that is independent of $n$, i.e., 
\begin{eqnarray}
&&R^{sp(\pm)}(E)=\frac{c}{4}|\mathcal M|^2\mathcal{D}_{\gamma}^{0}(E) \nonumber \\
 && \times \int_0^{\Delta E} \mathcal{D}_e(E')\mathcal{D}_h(\Delta E-E') n^e_{E'} n^h_{\Delta E-E'} {\rm d}E'. \quad \label{eq:Rsp}
\end{eqnarray}

Under the same approximation using the constant transition matrix element $\mathcal{M}$, the absorption coefficient $\alpha(E)$ is given microscopically by
\begin{eqnarray}
\alpha(E)&=& |\mathcal M|^2  \int_0^{\Delta E} \mathcal{D}_e(E')\mathcal{D}_h(\Delta E-E') \nonumber \\ 
&&\times (1-n^e_{E'}-n^h_{\Delta E-E'}) {\rm d}E'. \label{eq:alpha}
\end{eqnarray}

Using the homogeneous radiation rate and absorption coefficient, we consider continuity equations for the photon number current density inside the absorber.
Let $n^{\gamma (\pm)}_E$ and $j^{(\pm)}_E$ be the photon number density (per unit volume and per unit energy) and the photon number current density (per unit area, per unit time, and per unit energy) with energy $E$ propagating in the $\pm x$-directions.
Then, the continuity equations under the steady-state condition are given by 
\begin{eqnarray}
\partial_t n^{\rm \gamma (\pm )}_E=-\partial_x j^{(\pm)}_E(x)+ R^{sp(\pm)}(E)-\alpha(E)  j^{(\pm)}_E(x)=0,
\end{eqnarray} 
for $0<x<w$ (Fig.~\ref{fig5}).
The continuity equations under the appropriate boundary conditions,
\begin{eqnarray}
j^{(+)}_E(x \le 0)&=&j^{\rm sun}_E, \\
j^{(-)}_E(x \ge w)&=&0,
\end{eqnarray} 
give a solution for the output photon-number current density,
\begin{eqnarray}
j^{(+)}_E(x=w)&=&R^{sp(+)}(E)/\alpha(E), \\
j^{(-)}_E(x=0)&=&R^{sp(-)}(E)/\alpha(E),
\end{eqnarray}
where perfect absorption ($\alpha(E) w \gg 1$) and zero refrection at the front surface were assumed again.
From the results, the number of photons radiated out from the absorber per unit time ($\equiv {\rm d}N_\gamma^{Rad}/{\rm d}t$) through the front and back surfaces is
\begin{eqnarray}
\frac{{\rm d}N_\gamma^{Rad}}{{\rm d}t}&=&\mathcal{A} \int_{E_g}^{\infty}  (j^{(+)}_E(x=w)+j^{(-)}_E(x=0)) {\rm d}E \nonumber\\
&=& \frac{\mathcal{A}  c}{2 \pi ^2} \left( \frac{1}{\hbar c}\right)^3 \int_{E_g}^{\infty}
 \frac{E^2  \langle n^e n^h \rangle_{\Delta E}}{ \langle 1-n^e - n^h \rangle_{\Delta E} } {\rm d}E, \label{eq:GP} \qquad 
\end{eqnarray}
where Eq.~(\ref{eq:Rsp}) and Eq.~(\ref{eq:alpha}) were inserted, and 
\begin{eqnarray} 
&&\langle n^e n^h \rangle_{\Delta E} \nonumber \\
&&\equiv\frac{ \int_0^{\Delta E} \mathcal{D}_e(E')\mathcal{D}_h(\Delta E-E') n^e_{E'} n^h_{\Delta E-E'} {\rm d}E' }{ \int_0^{\Delta E}  \mathcal{D}_e(E')\mathcal{D}_h(\Delta E-E') {\rm d}E' },  \label{eq:NeNh} \\
&&\langle 1-n^e-n^h \rangle_{\Delta E} \nonumber \\
&&\equiv\frac{ \int_0^{\Delta E} \mathcal{D}_e(E')\mathcal{D}_h(\Delta E-E') (1-n^e_{E'}-n^h_{\Delta E-E'}) {\rm d}E' }{ \int_0^{\Delta E}  \mathcal{D}_e(E')\mathcal{D}_h(\Delta E-E') {\rm d}E' }. \nonumber \\ \label{eq:1-Ne-Nh}
\end{eqnarray} 
Equation Eq.~(\ref{eq:GP}) is a generalized Planck's law for indirect gap semiconductors. Given that the band filling effect can be neglected with $n^{e(h)}_{E_{e(h)}} \ll 1$, it is approximated using
\begin{eqnarray}
\frac{{\rm d}N_\gamma^{Rad}}{{\rm d}t}
&\approx& \frac{\mathcal{A}  c}{2 \pi ^2} \left( \frac{1}{\hbar c}\right)^3 \int_{E_g}^{\infty}
E^2  \langle n^e n^h \rangle_{\Delta E} {\rm d}E. \label{eq:GP2} \qquad 
\end{eqnarray}
Because the radiative loss of one photon is equal to the loss of one electron-hole pair, ${\rm d}N_\gamma^{Rad}/{\rm d}t$ can be related to $J^{e(h),{\rm rad}}_{E_e(h)}$.

{\it For indirect gap semiconductor absorbers ---} part of ${\rm d}N_\gamma^{Rad}/{\rm d}t$ in Eq.~(\ref{eq:GP2}) with Eq.~(\ref{eq:NeNh}), which comes from recombination of the electrons (holes) in a small energy window, $E_{e(h)}<E'<E_{e(h)}+{\rm d}E_{e(h)}$, divided by the number of corresponding electron (hole) states, $\mathcal{V} \mathcal{D}^{e(h)}_{E_{e(h)}} {\rm d}E_{e(h)}$, gives the radiation loss rates for the microscopic carrier distribution functions. This produces the following expressions: 
\begin{eqnarray}
J^{e,{\rm rad}}_{E_e}&=&\frac{c}{2 \pi ^2 w} \left( \frac{1}{\hbar c}\right)^3 \label{eq:Jerad-indirect} \\
 &\times & \int_{0}^{\infty}  \frac{(E_g+E_e+E_h)^2 \sqrt{E_h} \times n^e_{E_e} n^h_{E_h} }{(\pi/8) d_e  (E_e+E_h)^2} {\rm d}E_h, \nonumber \\ 
J^{h,{\rm rad}}_{E_h}&=&\frac{c}{2 \pi ^2 w} \left( \frac{1}{\hbar c}\right)^3 \label{eq:Jhrad-indirect}  \\
& \times & \int_{0}^{\infty}  \frac{(E_g+E_e+E_h)^2 \sqrt{E_e} \times n^e_{E_e} n^h_{E_h} }{(\pi/8)  d_h  (E_e+E_h)^2} {\rm d}E_e. \nonumber
\end{eqnarray}

{\it For direct gap semiconductor absorbers ---} taking the momentum conservation discussed in Sec.\ref{Jsun} into account, the radiation loss rates for the microscopic carrier distribution functions are obtained using a similar analysis. Here we simply show the final results: 
\begin{eqnarray}
J^{e,{\rm rad}}_{E_e}&=&\frac{c}{2 \pi ^2 w} \left( \frac{1}{\hbar c}\right)^3    \frac{(E_g+E_e+E_h)^2}{d_e \sqrt{E_e}} \label{eq:Jerad-direct}  \\
 &\times & \left.\left( 1+\frac{m_e^\ast}{m_h^\ast}\right) \frac{n^e_{E_e} n^h_{E_h}}{1-n^e_{E_e} - n^h_{E_h}} \right|_{E_h=\frac{m_e^\ast}{m_h^\ast}E_e} , \nonumber \\ 
J^{h,{\rm rad}}_{E_h}&=&\frac{c}{2 \pi ^2 w} \left( \frac{1}{\hbar c}\right)^3    \frac{(E_g+E_e+E_h)^2}{d_h \sqrt{E_h}} \label{eq:Jhrad-direct}  \\
 &\times & \left.\left( 1+\frac{m_h^\ast}{m_e^\ast}\right) \frac{n^e_{E_e} n^h_{E_h}}{1-n^e_{E_e} - n^h_{E_h}} \right|_{E_e=\frac{m_h^\ast}{m_e^\ast}E_h}. \nonumber 
\end{eqnarray}
Insertion of $1-n^e_{E_e} - n^h_{E_h}\approx 1$ into the denominators in Eq.~(\ref{eq:Jerad-direct}) and Eq.~(\ref{eq:Jhrad-direct}) gives approximate expressions for direct gap semiconductors that correspond to Eq.~(\ref{eq:Jerad-indirect}) and Eq.~(\ref{eq:Jhrad-indirect}) for indirect gap semiconductors. 

\subsection{\label{Jout} Carrier extraction rate: $J^{e(h),{\rm out}}_{E_{e(h)}}$}
The relaxation dynamics of a system of interest caused by weak interaction with the environment can be described using a standard approach that is widely used in studies of open quantum systems~\cite{Carmichael, Breuer}. The carrier extraction rate, $J^{e(h),{\rm out}}_{E_{e(h)}}$, is also obtained using a similar approach.
First, as shown in Fig.~\ref{fig1}, complete solar cell systems are divided into four parts: the main electron-hole systems in the absorber (Carrier System (sys) ), the electron and hole reservoirs (Bath 1 and Bath 2), and the phonon reservoir (Bath 3). 
Therefore, the noninteracting Hamiltonian for the whole system, $H_0$, can be given as the sum of four parts, i.e., $H_0=H_{0, {\rm sys}}+H_{0, {\rm Bath 1}}+H_{0, {\rm Bath 2}}+H_{0, {\rm Bath 3}}$, where
\begin{eqnarray}
&&H_{0, {\rm sys}}=\sum_{k}\left( (E_g+ E_e(k) ) e_k^\dagger e_k + E_h(k) h_k^\dagger h_k \right), \qquad \\
&&H_{0, {\rm Bath 1}}= \sum_{k'} \epsilon^c_{k'} c_{k'}^\dagger c_{k'}, \\
&&H_{0, {\rm Bath 2}}= \sum_{k'} \epsilon^d_{k'} d_{k'}^\dagger d_{k'}, \\
&&H_{0, {\rm Bath 3}}= \sum_{q} E^{ph}_{q} b_{q}^\dagger b_{q}.
\end{eqnarray}
Here, $e_k$ and $h_k$ are fermionic annihilation operators that are defined using the anticommutation relations, $[e_{k},e_{k'}^\dagger ]_+ = [h_{k},h_{k'}^\dagger ]_+=\delta_{k,k'}$ ($[X,Y]_+ \equiv XY+YX$), of the electrons and holes in the carrier system (absorber), with momentum $k$ and energies of $E_g+ E_e(k)$ and $E_h(k)$, respectively.
$c_{k'}$ and $d_{k'}$ represent the fermionic annihilation operators when defined using the anticommutation relations $[c_{k},c_{k'}^\dagger ]_+=[d_{k},d_{k'}^\dagger ]_+=\delta_{k,k'}$ for the electrons in Bath 1 and the holes in Bath 2 with momentum $k'$, and energies of $\epsilon^c_{k'} $ and $\epsilon^d_{k'} $, respectively.
$b_q$ represents a bosonic annihilation operator, which is defined using the commutation relation $[b_{q},b_{q'}^\dagger ]_-=\delta_{q,q'}$ ($[X,Y]_- \equiv XY-YX$) for the phonons in Bath 3 (the crystal lattice in the absorber) with momentum $q$ and energy $E^{ph}_{q} $.
Under the assumption of weak carrier system interaction with the environments (Bath 1 + Bath 2 + Bath 3),
the density matrix of the whole system $\rho$ can be approximated using a product of the matrices for the subsystems:
\begin{eqnarray}
\rho=\rho_{\rm sys} \otimes \rho_{\rm Bath 1} \otimes \rho_{\rm Bath 2} \otimes \rho_{\rm Bath 3}. \label{eq:DM}
\end{eqnarray} 
With this density matrix, the quantum and statistical average for any physical quantity $O$ is given by $\langle O \rangle={\rm Tr}(O \rho)$. 
Here, the density matrix for the Carrier System, denoted by $\rho_{\rm sys}$, is the matrix of interest and will be determined using the von Neumann equation~\cite{Breuer}. Additionally, the density matrices for the environments are assumed to be
\begin{eqnarray}
&&\rho_{\rm Bath 1}= \exp \left(-\beta_c (H_{0, {\rm Bath 1}}-\mu_c N_c)\right)/Z_{\rm Bath 1}, \qquad \\
&&\rho_{\rm Bath 2}= \exp \left(-\beta_c (H_{0, {\rm Bath 2}}-(-\mu_v) N_d)\right)/Z_{\rm Bath 2}, \qquad\\
&&\rho_{\rm Bath 3}= \exp \left(-\beta_{\rm ph} H_{0, {\rm Bath 3}} \right)/Z_{\rm Bath 3}, \qquad
\end{eqnarray} 
which represent the matrices in their thermal equilibrium states, e.g., with temperature $T_c$ and electron chemical potential $\mu_c$ for Bath 1, $T_c$ and hole chemical potential $\mu_h(=-\mu_v)$ for Bath 2, and phonon temperature $T_{\rm ph}$ and a chemical potential of zero for Bath 3. The $\beta$s represent inverse temperatures.
$N_c=\sum_{k'} c_{k'}^\dagger c_{k'}$ and $N_d=\sum_{k'} d_{k'}^\dagger d_{k'}$ represent the total numbers of carriers in Bath 1 and Bath 2, respectively, and the $Z$s are normalization factors used to ensure that
\begin{eqnarray}
{\rm Tr}(\rho_{\rm Bath 1})={\rm Tr}(\rho_{\rm Bath 2})={\rm Tr}(\rho_{\rm Bath 3})=1.
\end{eqnarray} 
Using the density matrix, the distribution functions for the carriers in the absorber are defined as
\begin{eqnarray}
&& n^e_k = n^e_{E_e(=E_e(k))} \equiv \langle e_k^\dagger e_k \rangle, \\
&& n^h_k = n^h_{E_h(=E_h(k))} \equiv \langle h_k^\dagger h_k \rangle,
\end{eqnarray}
while those for the particles in the baths are given by 
\begin{eqnarray}
\langle c^\dagger_{k'} c_{k'} \rangle &=&f^F_{\mu_c,\beta_c}(\epsilon^c_{k'}),\\
\langle d^\dagger_{k'} d_{k'} \rangle &=&f^F_{-\mu_v,\beta_c}(\epsilon^v_{k'}) \\
\langle b^\dagger_q b_q \rangle &=&f^B_{0,\beta_{\rm ph}}(E^{ph}_q),
\end{eqnarray}
where $f^F_{\mu,\beta}(E)(\equiv 1/(e^{\beta (E-\mu)}+1))$ and $f^B_{\mu,\beta}(E)(\equiv 1/(e^{\beta (E-\mu)}-1))$ are the Fermi-Dirac and Bose-Einstein distribution functions, respectively, with inverse temperature $\beta$ and chemical potential $\mu$. 

The electron extraction rate appears as a perturbation expansion to the kinetic motion in the equations for the distribution functions of the second-order with respect to the weak interaction between the Carrier System and Bath 1, $H'=H_{{\rm sys}-{\rm Bath 1}}$, whereas the interaction Hamiltonian is given using the form
\begin{eqnarray}
H_{{\rm sys}-{\rm Bath 1}}=\sum_{k,k'} (T^e_{k,k'} e_k c_{k'}^\dagger + (T^e_{k,k'})^\ast c_{k'} e_k^\dagger ). \label{eq:Ht}
\end{eqnarray}
The coupling parameter $T^e_{k,k'}$ represents the tunneling amplitude of an electron passing from the absorber through the tunnel barrier to the electron reservoir. 
Therefore, $T^e_{k,k'}$, is a function of the overlap integral of the wavefunctions in the absorber and the reservoir, i.e., it is a function of $k$ and $k'$, the height and width of the tunnel barrier, and the absorber thickness.

Switching to the interaction picture, where $O_I (t)\equiv e^{ i(H_0/\hbar) t}Oe^{ -i(H_0/\hbar) t}$, the von Neumann equation for the density matrix is given as
\begin{eqnarray}
\frac{d}{dt}\rho_I(t)=\frac{1}{i \hbar}[H_I'(t),\rho_I(t)]_-, \label{eq:vN0}
\end{eqnarray}
where 
\begin{eqnarray}
H_I'(t)=\sum_{k,k'} (T^e_{k,k'} e_k c_{k'}^\dagger e^{-i(E_g+E_e-\epsilon_{k'}^c)t/\hbar }+ {\rm h. c.}) 
\end{eqnarray}
is the interaction Hamiltonian, given by $H'(=H_{{\rm sys}-{\rm Bath 1}})$, in the interaction picture.
Successive iterations and time integration of Eq.~(\ref{eq:vN0}) gives
\begin{eqnarray}
\frac{d}{dt}\rho_I&=&\left(\frac{1}{i\hbar}\right)^2 \int_0^\infty [H_I'(t),[H_I'(t-\tau),\rho_I(t-\tau)]_-]_- {\rm d}\tau \nonumber \\
&\sim& \left(\frac{1}{i\hbar}\right)^2 \int_0^\infty [H_I'(t),[H_I'(t-\tau),\rho_I(t)]_-]_- {\rm d}\tau,  \label{eq:QME}
\end{eqnarray}
where the initial time contribution from $t=-\infty$ is neglected in the first equation, and a Markov approximation, under the assumption that the main system dynamics are sufficiently slow when compared with the memory time in the environments, is used in the second equation~\cite{Carmichael, Breuer}.  
Because we are considering steady-state operation of the solar cells, the Markov approximation can be used safely. 

Using Eq.~(\ref{eq:QME}), the extraction rate for an electron with kinetic energy $E_e(=E_e(k))$ for momentum $k$ is given by
\begin{eqnarray}
&&-J^{e,{\rm out}}_{E_{e}}= 
\frac{d}{dt}\langle e_k^\dagger e_k \rangle = {\rm Tr} \left( e_k^\dagger e_k \dot{\rho}_I(t)\right)   \label{eq:tunnel1} \\
&&= \left(\frac{1}{i\hbar}\right)^2 \int_0^\infty {\rm Tr} \left( e_k^\dagger e_k  [H_I'(t),[H_I'(t-\tau),\rho_I(t)]_-]_- \right) {\rm d}\tau \nonumber
\end{eqnarray}
Using the cyclic property of the trace, where ${\rm Tr}(XYZ)={\rm Tr}(ZXY)$, the integrand on the right-hand side of the third equation is given explicitly by
\begin{eqnarray}
&& {\rm Tr}\left(
\left( e_k^\dagger e_k H_I'(t)H_I'(t-\tau)  -H_I'(t-\tau)H_I'(t)e_k^\dagger e_k \right. \right.  \label{eq:Jeout-int} \\ 
&& \left. \left. - H_I'(t-\tau)e_k^\dagger e_k H_I'(t) + H_I'(t)e_k^\dagger e_k H_I'(t-\tau) 
\right) \rho_I(t)
 \right). \nonumber
\end{eqnarray}
Considering the assumption in Eq.~(\ref{eq:DM}), the trace for the whole system can be provided by successive partial traces of the subsystems. By retaining only the terms that do not vanish after the trace is taken, Eq.~(\ref{eq:Jeout-int}) can be rewritten as
\begin{eqnarray}
&&\sum_{k'}|T^e_{k.k'}|^2  (e^{i(E_g+E_e-\epsilon_{k'}^c) \tau/\hbar}-e^{-i(E_g+E_e-\epsilon_{k'}^c) \tau/\hbar }) \nonumber \\ &&
\times {\rm Tr}\left(
( e_k^\dagger e_k e^\dagger_{k} e_{k} c_{k'}c_{k'}^\dagger  -e_k e_k^\dagger e_k e_k^\dagger c_{k'}^\dagger c_{k'})  \rho_I (t) \right) \nonumber \\
&=&\sum_{k'}|T^e_{k.k'}|^2  (e^{i(E_g+E_e-\epsilon_{k'}^c) \tau/\hbar}-e^{-i(E_g+E_e-\epsilon_{k'}^c) \tau/\hbar })  \nonumber \\ 
&&\times \left(n_{E_e}^e (1-f^{F}_{\mu_c, \beta_c}(\epsilon_{k'}^c)) -(1-n_{E_e}^e) f^{F}_{\mu_c, \beta_c}(\epsilon_{k'}^c) \right) . \qquad  \label{eq:tunnel2}
\end{eqnarray}  
By inserting Eq.~(\ref{eq:tunnel2}) into Eq.~(\ref{eq:tunnel1}) and performing the integration with respect to time, we obtain
\begin{eqnarray}
J^{e,{\rm out}}_{E_{e}}&=& \frac{2 \pi }{\hbar}\sum_{k'}|T^e_{k.k'}|^2  \delta(E_g+E_e-\epsilon_{k'}^c) \nonumber \\
&& \times \left( n^e_{E_e}-f^{F}_{\mu_c, \beta_c}(E_g+E_e) \right) .
\end{eqnarray}  
We therefore derive a simple expression for the extraction rate:
\begin{eqnarray}
J^{e,{\rm out}}_{E_{e}}&=&\frac{1}{\tau_{\rm out}^e} \left( n^e_{E_e}-f^{F}_{\mu_c, \beta_c}(E_g+E_e) \right), \label{eq:Jeout1}
\end{eqnarray}  
with extraction time $\tau_{\rm out}^e$ that is defined as
\begin{eqnarray}
(\tau_{\rm out}^e)^{-1}= \left. \frac{2 \pi }{\hbar}|T^e(E)|^2 \mathcal{D}_c(E) \right|_{E=E_g+E_e}, \label{eq:tout-e}
\end{eqnarray}
where $\mathcal{D}_c(E) (= \sum_{k'} \delta(E-\epsilon_{k'}^c))$ is the density of states in Bath 1, and 
\begin{eqnarray}
|T^e(E)|^2 &=&\frac{\sum_{k'}|T^e_{k, k'}|^2  \delta(E-\epsilon_{k'}^c)}{\sum_{k'}  \delta(E-\epsilon_{k'}^c)},
\end{eqnarray}
represents the strength of tunneling coupling when averaged over the states in Bath 1 at energy $E$. 
While $\tau_{\rm out}^e$ is dependent on the energy of the carriers, we consider it to be a constant parameter in the following analysis for simplicity. 
In this sense, the carrier extraction time used here is an effective parameter representative for all the electrons tunneling between the absorber and the electrode. 

The same argument is also applicable to the hole extraction rate when using $H'=H_{\rm sys-Bath2}=\sum_{k,k'} (T^h_{k,k'} h_k d_{k'}^\dagger +{\rm h.c.} )$. 
We therefore have a similar expression for holes:
\begin{eqnarray}
J^{h,{\rm out}}_{E_{h}}&=&\frac{1}{\tau^h_{\rm out}} \left( n^h_{E_h}-f^{F}_{-\mu_v, \beta_c}(E_h) \right),
\label{eq:Jhout1}
\end{eqnarray}  
with a hole extraction time of
\begin{eqnarray}
(\tau_{\rm out}^h)^{-1}&=&\frac{2 \pi }{\hbar}\sum_{k'}|T_{k.k'}^h|^2  \delta(E_h-\epsilon_{k'}^d) 
\nonumber \\
&\equiv&\left. \frac{2 \pi }{\hbar}|T^h(E)|^2 \mathcal{D}_d(E)\right|_{E=E_h}, \label{eq:tout-h}
\end{eqnarray}
where $|T^h(E)|^2$ is an effective tunneling probability for a hole tunneling from the absorber to Bath 2, and $\mathcal{D}_d(E)(= \sum_{k'}  \delta(E-\epsilon_{k'}^d))$ is the density of states in Bath 2 at energy $E$. 
In the following, we also assume the simplest case, i.e., where $\tau^e_{\rm out}=\tau^h_{\rm out} \equiv \tau_{\rm out}$, which will not limit the generality of the main conclusion.

As mentioned earlier, $\tau_{\rm out}$ should be regarded as the effective time scale used for carrier extraction.
In this sense, however, it could also be used to parametrize the time between photogeneration and extraction of the carriers, which may be required for other reasons; e.g., when photogeneration occurs at the center of absorber, $\tau_{\rm out}$ cannot be less than the time delay given by the distance from the point of generation to the contacts divided by the average carrier velocity.
Such time delays would be important in thicker solar cells and appear to be critical in solar cells using nanocrystals, organic solar cells~\cite{Tang, Heeger} (including dye-sensitised~\cite{Gratzel}), and perovskite solar cells~\cite{Miyasaka}, which have low carrier mobilities caused by disorders and the Frenkel-like localization~\cite{Frenkel, MarkFox} of excitons. 
$\tau_{\rm out}$ could be measured using specific characterization methods that are suitable for each system, e.g., by optical characterization of the lifetimes of photo-generated carriers under short-circuit conditions and by transient measurement of the photocurrents~\cite{Ishii, Koster}.

The description based on the tunnel Hamiltonian given in Eq.~(\ref{eq:Ht}) simply neglects the voltage drop at the contact. 
Ohmic contacts with energy losses at the contacts are outside the scope of this paper.   

\subsection{\label{Jph} Phonon scattering (thermalization) rate: $\left.\frac{d}{dt}n^{e(h)}_{E_{e(h)}}\right|_{\rm phonon}$}
The phonon scattering (thermalization) rate in the rate equation for the microscopic carrier distribution functions, given by $\left.\frac{d}{dt}n^{e(h)}_{E_{e(h)}}\right|_{\rm phonon}$, can be obtained using an analysis similar to those presented in Sec.~\ref{Jout}.
In this case, interactions between the carriers and the phonons are considered using the perturbation Hamiltonian $H'=H_{\rm sys-Bath3}$, which is given by
\begin{eqnarray} 
&&H_{\rm sys-Bath3} =\sum_{q,k}g^c_{q} (b_q+b_{-q}^\dagger)(e_{k+q}^\dagger e_k
 +{\rm h.c.}) \nonumber \\
&&+\sum_{q,k}g^v_{q} (b_q+b_{-q}^\dagger)(h_{k+q}^\dagger h_k
 +{\rm h.c.}),
\end{eqnarray}
where $g^c_{q}$ and $g^v_{q}$ are the electron-phonon coupling constants for the bottom-conduction-band and top-valence-band electrons in the absorber, respectively. The $q$-dependence of the coupling constants is dependent on the types of phonons involved. For an order of magnitude-level estimate of the thermalization rate, the LA phonons that originate from the deformation potential are considered for Si absorbers:
\begin{eqnarray}
g^{c(v)}_{q}=a_{{\rm def}, c(v)} \sqrt{\frac{\hbar q}{2\mathcal{V} v_A \rho_A }}, \label{eq:gLA}
\end{eqnarray}
where $a_{{\rm def}, c(v)} $, $v_A$, and $\rho_A$ are the deformation potential for the conduction (valence) electrons, the phonon velocity, and the mass density in the absorber, respectively. A realistic estimate requires inclusion of the scattering caused by the other phonon modes, i.e., the transverse acoustic (TA), transverse optical (TO), and LO modes, and the intra-valley scattering (within the degenerate bands)~\cite{Suzuki}, based on realistic electron and phonon band structures~\cite{Cardona}, which is far beyond the scope of this work. 

The interaction Hamiltonian in the interaction picture is
\begin{eqnarray}
&&H'_I(t)=\sum_{q,k}\left( g^c_{q} e_{k+q}^\dagger e_k b_q e^{i(E_e(k+q)-E_e(k)-E^{ph}_q)t} +{\rm h.c.} \right) \nonumber  \\
&&  + \sum_{q,k} \left( g^c_{q} e_{k+q}^\dagger e_k b_{-q}^\dagger  e^{i(E_e(k+q)-E_e(k)+E^{ph}_{q})t} +{\rm h.c.} \right)
 \nonumber \\
&&  + \sum_{q,k} \left( g^v_{q} h_{k+q}^\dagger h_k b_{q}  e^{i(E_h(k+q)-E_h(k)-E^{ph}_{q})t} +{\rm h.c.} \right) \label{eq:HI-phonon} \\
&&\ +\sum_{q,k} \left( g^v_{q} h_{k+q}^\dagger h_k b_{-q}^\dagger e^{i(E_h(k+q)-E_h(k)+E^{ph}_{q})t} +{\rm h.c.} \right). \nonumber
\end{eqnarray}
Insertion of Eq.~(\ref{eq:HI-phonon}) into ${\rm Tr}(e^\dagger_k e_k \dot{\rho}_I(t) )$ with the second-order Born-Markov approximation given in Eq.~(\ref{eq:QME}) and performing a time integration, we obtain the phonon scattering rates as follows:
\begin{eqnarray}
&&\left.\frac{d}{dt}n^e_{E_e}\right|_{\rm phonon} =-\frac{2 \pi}{\hbar} \sum_q  (g^c_{q})^2  \nonumber \\
 && \times \delta \bigl(E_{e}(k)-E_e(k-q)-E_q^{ph} \bigr)  \Bigl( n^e_{k}(1-n_{k-q}^e) \nonumber \\
&&   \left. \times (f^B_{0,\beta_{\rm ph}}(E^{ph}_q)+1)-n_{k-q}^e (1-n_{k}^e)  f^B_{0,\beta_{\rm ph}}(E^{ph}_q) \right) \nonumber \\
 &&+\frac{2 \pi}{\hbar} \sum_q(g^c_{q})^2 \delta  \bigl( E_e(k+q)-E_e(k)-E_q^{ph}  \bigr)   \nonumber \\
&&  \times \Bigl( n^e_{k+q}(1-n_{k}^e)(f^B_{0,\beta_{\rm ph}}(E^{ph}_q)+1) \nonumber \\
&& \qquad \qquad \qquad -n_{k}^e (1-n_{k+q}^e)  f^B_{0,\beta_{\rm ph}}(E^{ph}_q) \Bigr),   \label{eq:Je-themal0} \\
&&\left.\frac{d}{dt}n^h_{E_h}\right|_{\rm phonon} =-\frac{2 \pi}{\hbar} \sum_q  (g^v_{q})^2  \nonumber \\
 && \times \delta \bigl(E_{h}(k)-E_h(k-q)-E_q^{ph} \bigr)  \Bigl( n^h_{k}(1-n_{k-q}^h) \nonumber \\
&&   \left. \times (f^B_{0,\beta_{\rm ph}}(E^{ph}_q)+1)-n_{k-q}^h (1-n_{k}^h)  f^B_{0,\beta_{\rm ph}}(E^{ph}_q) \right) \nonumber \\
 &&+\frac{2 \pi}{\hbar} \sum_q(g^v_{q})^2 \delta  \bigl( E_h(k+q)-E_h(k)-E_q^{ph}  \bigr)   \nonumber \\
&&  \times \Bigl( n^h_{k+q}(1-n_{k}^h)(f^B_{0,\beta_{\rm ph}}(E^{ph}_q)+1) \nonumber \\
&& \qquad \qquad \qquad -n_{k}^h (1-n_{k+q}^h)  f^B_{0,\beta_{\rm ph}}(E^{ph}_q) \Bigr).   \label{eq:Jh-themal0}
\end{eqnarray}
Scattering rates in this form are equivalent to those obtained using Fermi's golden rule calculation.
The expression above includes a momentum representation of the carrier distribution functions that can be further stated using simpler expressions in the energy representation.
First, we transform the $q$-summation into an integration over polar coordinates in the form $\sum_q=\frac{\mathcal{V}}{(2 \pi)^3}\int_0^\infty 2 \pi |q|^2 {\rm d}|q| \int_{-1}^{1} {\rm d} (\cos{\theta})$, where $\theta$ is the angle between $k$ and $q$. 
The angular integration is then performed using the dispersion relation $E_q^{ph}=\hbar v_A |q|$, the coupling constants for LA phonons in Eq.~(\ref{eq:gLA}), and
\begin{eqnarray}
E_{e(h)}(k)-E_{e(h)}(k-q) =\frac{\hbar^2(2|q||k|\cos{\theta}-|q|^2)}{2m^\ast_{e(h)}}, \\
E_{e(h)}(k+q)-E_{e(h)}(k) =\frac{\hbar^2(2|q||k|\cos{\theta}+|q|^2)}{2m^\ast_{e(h)}}.
\end{eqnarray}
By making a change in the coordinates, where $|q|=\epsilon/(\hbar v_A)$, we finally obtain
\begin{eqnarray}
&&\left.\frac{d}{dt}n^{e(h)}_{E_{e(h)}}\right|_{\rm phonon}  \nonumber \\
&=& 
-C_{ph}^{e(h)}  \int_0^{\epsilon_{\rm cut}^{e(h),-}} \frac{\epsilon^2d\epsilon}{\sqrt{E_{e(h)}}}
 \biggl( n^{e(h)}_{E_{e(h)}}(1-n_{E_{e(h)}-\epsilon}^{e(h)}) \nonumber \\
&\times&  \bigl( 1+f^B_{0, \beta_{\rm ph}} ({\epsilon})\bigr)-n_{E_{e(h)}-\epsilon}^{e(h)} (1-n^{e(h)}_{E_{e(h)}})  f^B_{0, \beta_{\rm ph}} ({\epsilon}) \biggr) 
 \nonumber \\
&+&
C_{ph}^{e(h)}  \int_0^{\epsilon_{\rm cut}^{e(h),+}} \frac{\epsilon^2d\epsilon}{\sqrt{E_{e(h)}}}
 \biggl( n^{e(h)}_{E_{e(h)}+\epsilon}   (1-n^{e(h)}_{E_{e(h)}})  \nonumber \\
&\times&  \bigl(1+ f^B_{0, \beta_{\rm ph}} ({\epsilon})\bigr)-n_{E_{e(h)}}^{e(h)} (1-n_{E_{e(h)}+\epsilon}^{e(h)}) f^B_{0, \beta_{\rm ph}} ({\epsilon}) \biggr) .\qquad \label{eq:Jeh-themal1}
\end{eqnarray}
Terms proportional to $f^B_{0, \beta_{\rm ph}} ({\epsilon})$ and $1+f^B_{0, \beta_{\rm ph}} ({\epsilon})$ represent the scattering rates for phonon absorption and emission, respectively. 
Here, we newly defined the coefficient $C_{ph}^{e(h)}\equiv \frac{ a_{\rm def,c(v)}^2 \sqrt{m_{e(h)}^\ast/2 }}{4 \pi \hbar^4 v_A^4 \rho_A }$ and the $E_{e(h)}$-dependent cutoff energies as follows:
\begin{eqnarray}
&&\epsilon_{\rm cut}^{e(h),-}  \equiv \min \biggl(  E_{e(h)}, \epsilon_{\rm cut}^{ph}, \nonumber \\
&& \qquad \qquad  2v_A \bigl( \sqrt{2m_{e(h)}^\ast E_{e(h)}}-m_{e(h)}^\ast v_A \bigr) \biggr), \label{eq:EphCut-} \\
&& \epsilon_{\rm cut}^{e(h),+}\equiv \min \biggl(  \epsilon_{\rm cut}^{ph},  2v_A \bigl( \sqrt{2m_{e(h)}^\ast E_{e(h)}}+m_{e(h)}^\ast v_A \bigr) \biggr), \qquad \label{eq:EphCut+}
\end{eqnarray}
where $\epsilon_{\rm cut}^{ph}$ is the Debye cutoff energy for the LA phonons ($\sim50$ meV for Si). The $E_{e(h)}$ dependences in Eq.~(\ref{eq:EphCut-}) and Eq.~(\ref{eq:EphCut+}) stem from the condition that the arguments in the delta functions in Eq.~(\ref{eq:Je-themal0}) and Eq.~(\ref{eq:Jh-themal0}) are zero, i.e., from the requirements for energy and momentum conservation during the carrier-phonon scattering processes.
A situation also occurs in which the cutoff energy $\epsilon_{\rm cut}^{e(h),-}$ becomes negative. 
This occurs when $\sqrt{2m_{e(h)}^\ast E_{e(h)}}-m_{e(h)}^\ast v_A <0 $ for small $E_{e(h)}$, i.e., when $E_{e(h)}< \frac{1}{2}m_{e(h)}^\ast v_A^2 \equiv E^{\rm PB}_{e(h)}$.
In this case, carriers with $E_{e(h)}< E^{\rm PB}_{e(h)}$ cannot lose energy via phonon emissions (i.e., carrier cooling does not occur), even at the zero temperature of the lattice ($T_{\rm ph}=0$), which is prohibited by the conservation law. 
However, this
 Effect, which is called the phonon bottleneck effect, is normally negligible because the threshold energy $E^{PB}_{e(h)}$, known as the phonon bottleneck energy, is very low (e.g., $v_A =10^4$ m/sec, $m_e^\ast/m_e=1.08$, and $m_h^\ast/m_e=0.55$ give $E^{\rm PB}_{e}=0.307$ meV and $E^{\rm PB}_{h}=0.156$ meV in the model for Si). 
We have already implicitly assumed that $E_{e(h)}> E^{\rm PB}_{e(h)}$ in Eq.~(\ref{eq:Jeh-themal1}) (which sets $\epsilon_{\rm cut}^{e(h),-}>0$ and the lower domain boundary of the latter integration to zero).

\begin{figure}[!t]
\centering
\includegraphics[width=0.48\textwidth]{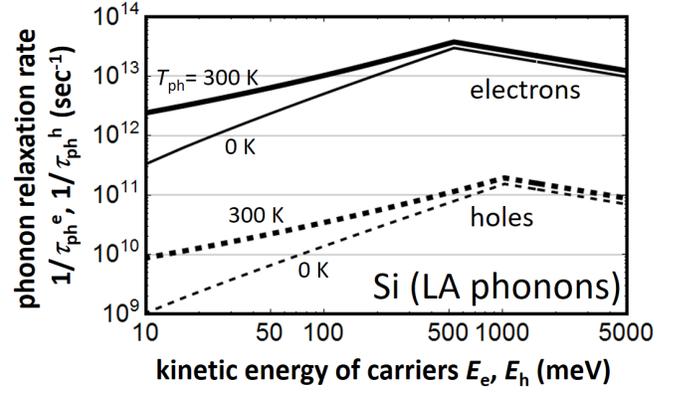}
\caption{Estimation of phonon relaxation rate of carriers via the LA mode in Si as estimated from Eq.~(\ref{eq:PhononScatteringRate}) as a function of kinetic energy for the electrons (solid) and holes (dashed) at two different lattice temperatures: $T_{\rm ph}=300$ K (thick) and 0 K (thin). The following parameters are used for Si: $v_A=10^4$ m/s, $m_e^\ast/m_e=1.08$, $m_h^\ast/m_e=0.55$, $\epsilon_{\rm cut}^{ph}=50$ meV, $a_{\rm def,c}=10$ eV, $a_{\rm def,v}=1$ eV, and $\rho_A=2.3$ g/cm$^3$. }  
\label{fig6}  
\end{figure}

The time scale for thermalization of the photogenerated carriers in the absorber can be estimated from Eq.~(\ref{eq:Jeh-themal1}).
Consider the case where electrons (holes) with energy $E_{e(h)}$ are generated at an initial time $t=0$ under illumination by a narrow-band photon source at the corresponding energy.
In this case, $n^{e(h)}_{E_{e(h)}}\ne 0$ and $n^{e(h)}_{E \ne E_{e(h)}} =0$ at $t=0$. When this condition is inserted into Eq.~(\ref{eq:Jeh-themal1}), the initial population dynamics are given by
\begin{eqnarray}
\frac{d}{dt}n^{e(h)}_{E_{e(h)}}=-\frac{1}{\tau^{e(h)}_{\rm ph}}n^{e(h)}_{E_{e(h)}},
\end{eqnarray}
where the relaxation time $\tau^{e(h)}_{\rm ph}$ is given by
\begin{eqnarray}
\tau^{e(h)}_{\rm ph}&=&C_{ph}^{e(h)}  \biggl(  \int_0^{\epsilon_{\rm cut}^{e(h),-}} \frac{\epsilon^2 \bigl(1+ f^B_{0, \beta_{\rm ph}} ({\epsilon})  \bigr)  }{\sqrt{E_{e(h)}}}  d\epsilon \nonumber \\
&&+ \int_0^{\epsilon_{\rm cut}^{e(h),+}} \frac{\epsilon^2   f^B_{0, \beta_{\rm ph}} ({\epsilon})}{\sqrt{E_{e(h)}}}   d\epsilon \biggr).  \label{eq:PhononScatteringRate}
\end{eqnarray}
At the zero temperature of the lattice, insertion of $f_{0,\beta_{\rm ph}=+\infty}^{B}(\epsilon)=0$ into Eq.~(\ref{eq:PhononScatteringRate}) gives the simple expression
\begin{eqnarray}
\frac{1}{\tau^{e(h)}_{\rm ph}}=C_{ph}^{e(h)} \frac{(\epsilon_{\rm cut}^{e(h),-})^3}{3 \sqrt{E_e(h)}}. \label{eq:PhononScatteringRate0}
\end{eqnarray}
Of course, the time constant $\tau^{e(h)}_{\rm ph}$ gives a timescale for the initial relaxation dynamics for such an ideal situation,
It therefore seems reasonable to consider that the carrier thermalization time in solar cells could also be estimated using $\tau^{e(h)}_{\rm ph}$ in Eq.~(\ref{eq:PhononScatteringRate}).  
Fig.~\ref{fig6} shows the phonon relaxation rate that was estimated for Si as a function of the kinetic energies of the carriers.
In our Si model, we found that the relaxation time ranges between $10^{-10}$ and $10^{-13.5}$ s in the relevant energy window for solar cells (as shown in Fig.~\ref{fig5} that corresponds to the bandwidth of the solar spectra, $k_B T_S$).
This result is consistent with the measured timescale for the carrier cooling process, which ranges from sub-picosecond to hundreds of picoseconds~\cite{Goldman, Sabbah, Suzuki} and also with the timescales in the literature~\cite{BookGreen}.  
A major difference (two orders of magnitude) between the results for electrons and holes originates from differences in the deformation potentials for the LA phonons.
The situation can therefore change if other phonon modes and the fast electron-hole equilibration that was discussed in \cite{Suzuki} are taken into account in the calculations. 
From this perspective, we should stress here again that the results in Fig.~\ref{fig6} were given for an order of magnitude-level estimation.

\subsection{\label{broadening} Effects of spectral broadening of the microscopic states}
In the previous four subsections (Sec.~\ref{Jsun}, Sec.~\ref{Jrad}, Sec.~\ref{Jout}, Sec.~\ref{Jph}), explicit forms of the rate equation for the microscopic carrier distribution function in both Eq.~(\ref{eq:e-balance}) and Eq.~(\ref{eq:h-balance}) appear to have been found.
Actually, as shown in the simulations below, direct use of the equations that have been derived thus far can be justified in most cases.
However, in certain situations, the equations should be modified slightly.
Modifications are required when the spectral broadening of the microscopic carrier states becomes important.
These situations occur when the carriers are far out of equilibrium.
The spectral broadening $\Gamma$ is taken into account in the NEGF formalism automatically as an imaginary part of the (retarded) self-energy ($\Gamma = -{\rm Im} \Sigma^{\rm r} $)~\cite{Martin-Schwinger, Kadanof-Baym, Keldysh}. 
Therefore, we also take the broadening effect into account in a satisfactory manner while keeping the calculations as simple as possible.

Several factors can broaden the spectra of the single particle states in many-particle systems.
One factor comes from Coulomb interaction between the carriers.
However, we consider the Coulomb interaction to have a minor or secondary effect on the properties of solar cells, which normally work at low carrier densities (as shown for Si solar cells in~\cite{Richter}), and we have already neglected to include it in our model.
Other factors occur because of interactions between the carrier system and the baths, or more explicitely, because of the carrier-phonon interaction and carrier extraction processes that occur in our model.
Spectral broadening of the electrons (holes) with $E_{e(h)}$, given by $\Gamma^{e(h)}_{E_{e(h)}} (\equiv -{\rm Im} \Sigma^{r} (E_{e(h)}))$, is related to the time constants $\tau_{\rm out}^{e(h)}(=\tau_{\rm out})$ and $ \tau_{\rm ph}^{e(h)}$, which we already determined in the preceding subsections:
\begin{eqnarray}
\Gamma^{e(h)}_{E_{e(h)}}=\frac{\hbar}{2 \tau_{\rm ph}^{e(h)}}+\frac{\hbar}{2 \tau_{\rm out}^{e(h)}}. \label{eq:Gamma}
\end{eqnarray}
Therefore, the broadening is given by $\hbar/(2 \tau_{\rm ph}^{e(h)})$ when $\tau_{\rm ph}^{e(h)} \ll \tau_{\rm out}^{e(h)}$ and $\hbar/(2 \tau_{\rm out}^{e(h)})$ when $\tau_{\rm out}^{e(h)} \ll \tau_{\rm ph}^{e(h)}$.

In the former case ($ \tau_{\rm ph}^{e(h)} \ll \tau_{\rm out}^{e(h)}$), it is natural to consider that the carriers must be fully relaxed to establish thermal equilibrium with the lattice in the steady state.
This assumption can be checked by solving the rate equations in Eq.~(\ref{eq:e-balance}) and Eq.~(\ref{eq:h-balance}) as follows.
When the thermalization term $\left.\frac{d}{dt}n^{e(h)}_{E_{e(h)}}\right|_{\rm phonon}$ dominates the rate equation, a steady state is achieved when all integrands in Eq.~(\ref{eq:Jeh-themal1}) disappear. 
This condition is fulfilled if
\begin{eqnarray}
&&
\left( \frac{n^{e(h)}_{E_{e(h)}+\epsilon} }{ 1-n^{e(h)}_{E_{e(h)}+\epsilon} } \right) 
\Bigg/ 
\left( \frac{n^{e(h)}_{E_{e(h)}} }{ 1-n^{e(h)}_{E_{e(h)}} } \right) \nonumber \\
&=&\frac{f^{B}_{0,\beta_{\rm ph}}(\epsilon)}{1+f^{B}_{0,\beta_{\rm ph}}(\epsilon)}
=\exp(-\beta_{\rm ph}\epsilon) \nonumber \\
&=& \exp(-\beta_{\rm ph}(E_{e(h)}+\epsilon)) /
\exp(-\beta_{\rm ph}E_{e(h)})
\end{eqnarray}
for every possible choice of $E_{e(h)}$ and $\epsilon$. 
This means that in the steady state, the distribution function is given by the Fermi-Dirac distribution, with some value for the chemical potential in the absorber $\mu^{\rm cell}_{e(h)}$:
\begin{eqnarray}
n^{e(h)}_{E_{e(h)}}=f^{F}_{\mu^{\rm cell}_{e(h)}, \beta_{\rm ph}}(E_{e(h)}). \label{eq:FDfuncPartialEquilibrium}
\end{eqnarray}
Therefore, these distribution functions are thermally distributed over the energy range $0<E_{e(h)}<k_B T_{\rm ph}$. 
In this case, inclusion of the broadening within the single particle spectra does not modify the distribution function as long as $\Gamma^{e(h)}_{E_{e(h)}}\left(=\hbar/(2\tau^{e(h)}_{\rm ph})\right)< k_B T_{\rm ph}$ (=26 meV for 300 K). 
The condition appears to be satisfied well when we consider that a phonon relaxation time of 1 ps corresponds to broadening of 0.33 meV. 
In this way, we confirm that the broadening effect is negligible for $ \tau_{\rm ph}^{e(h)} \ll \tau_{\rm out}^{e(h)}$.

The above consideration allows us to neglect broadening due to electron-phonon interactions in Eq.~(\ref{eq:Gamma}) over the entire range of $\tau_{\rm out}^{e(h)}$. We can therefore safely use the following approximation: 
\begin{eqnarray}
\Gamma^{e(h)}_{E_{e(h)}}=\frac{\hbar}{2 \tau_{\rm out}^{e(h)}}. \label{eq:Gamma2}
\end{eqnarray}
This approximation greatly simplifies the calculations because $\tau_{\rm out}^{e(h)}(=\tau_{\rm out})$ is a given parameter that is identical for every carrier state in our model, whereas $\tau^{e(h)}_{\rm ph}$, which should be defined precisely from Eq.~(\ref{eq:Jeh-themal1}), is a function of the carrier distribution functions that are to be determined finally. 
We therefore use an approximation based on use of Eq.~(\ref{eq:Gamma2}) for broadening of the single-particle states in this work.
We define the lineshape function of the states using $\mathcal{A}_\Gamma (x)$, which is a Gaussian function with a half width at half maximum that is equal to $\Gamma^{e(h)}_{E_{e(h)}}$. 
Using $\Gamma \equiv (\log{2})^{-1/2} \Gamma^{e(h)}_{E_{e(h)}}$, the function is given by 
\begin{eqnarray}
\mathcal{A}_\Gamma (x)=\sqrt{\frac{1}{\pi \Gamma^2}}\exp \left( -(x/\Gamma )^2\right).
\end{eqnarray}
A Lorentzian function is normally used for the lineshape of a single particle state (this also applies in the NEGF formalism~\cite{Datta}) rather than a Gaussian spectral profile because the Gaussian distribution reflects the statistical fluctuations of the system; it is not derived naturally for ideal models without structural imperfections.
We use the Gaussian function for a reason that is demonstrated later in the paper; however, it will not change our main conclusion. Using the line function, expressions for the generation and loss rates in the rate equation that has been derived thus far are modified as follows, where the modification becomes important when $\tau_{\rm out} \ll  \tau_{\rm ph}^{e(h)}$.

{\it For the generation rate $J^{e(h),{\rm sun}}_{E_{e(h)}}$} --- because the Sun's spectrum is much broader than the broadening of the states, $k_B T_S \gg \Gamma$, we can neglect the effects of broadening on $J^{e(h),{\rm sun}}_{E_{e(h)}}$. We therefore use Eq.~(\ref{eq:Je-sun-direct}) and Eq.~(\ref{eq:Jh-sun-direct}) for direct gap semiconductors and Eq.~(\ref{eq:Jerad-indirect}) and Eq.~(\ref{eq:Jhrad-indirect}) for direct gap semiconductors.

{\it For the radiative recombination loss rate $J^{e(h),{\rm rad}}_{E_{e(h)}}$} --- inclusion of the spectral broadening modifies the density of states of the carriers.
This effect replaces $\mathcal{D}_{e(h)}(E)$ with $\tilde{\mathcal{D}}_{e(h)}(E)$ in the analysis that was presented in Sec.~\ref{Jrad} (particularly in Eq.~(\ref{eq:Rsp}), Eq.~(\ref{eq:alpha}), Eq.~(\ref{eq:NeNh}), and Eq.~(\ref{eq:1-Ne-Nh})), where
\begin{eqnarray}
\tilde{\mathcal{D}}_{e(h)}(E)&\equiv& \left( \mathcal{D}_{e(h)} \ast \mathcal{A}_{\Gamma} \right)(E) \nonumber \\
&=& \int_0^\infty {\mathcal{D}}_{e(h)}(E')\mathcal{A}_{\Gamma}(E-E') {\rm d}E', \label{eq:Deh-renew}
\end{eqnarray}
is the density of states of the carriers convolved using the spectral function. For example, the denominator in both Eq.~(\ref{eq:NeNh}) and Eq.~(\ref{eq:1-Ne-Nh}) can be modified as follows: 
\begin{eqnarray}
&& \int_{-\infty}^{\infty}  \tilde{\mathcal{D}}_e(E')\tilde{\mathcal{D}}_h(\Delta E-E') {\rm d}E' \nonumber \\
&=& \int_{-\infty}^\infty {\rm d}E' \int_{0}^\infty {\rm d}E_e \int_{0}^\infty {\rm d}E_h  \mathcal{D}_e(E_e) \mathcal{A}_\Gamma (E'-E_e) \nonumber \\
&& \times  \mathcal{D}_h(E_h) \mathcal{A}_\Gamma (\Delta E-E'-E_h) \nonumber \\
&=& \int_{0}^\infty {\rm d}E_e \int_{0}^\infty {\rm d}E_h  \mathcal{D}_e(E_e)  \mathcal{D}_h(E_h) \nonumber \\
&& \qquad \qquad  \times  \mathcal{A}_{\sqrt{2}\Gamma} (\Delta E-E_e-E_h) .  \label{eq:Normalization-renew}
\end{eqnarray}
In the last equation, we used a general property of the convolution of Gaussian functions:
\begin{eqnarray}
\mathcal{A}_{\Gamma_1}\ast \mathcal{A}_{\Gamma_2}(E)&\equiv&
 \int_{-\infty}^\infty {\rm d}E'\mathcal{A}_{\Gamma_1}(E') \mathcal{A}_{\Gamma_2}(E-E') \nonumber \\
&=& \mathcal{A}_{\sqrt{\Gamma_1^2+\Gamma_2^2}}(E).
\end{eqnarray}
A similar modification was made to the numerator in Eq.~(\ref{eq:NeNh}) and Eq.~(\ref{eq:1-Ne-Nh}).
As a result, the recombination loss rates for the indirect transition are given by 
\begin{eqnarray}
&& J^{e, {\rm rad}}_{E_e}= \frac{1}{w} \times \frac{c}{2 \pi ^2} \left( \frac{1}{\hbar c}\right)^3 \times  \int_{0}^{\infty}  {\rm d}E \  E^2  \label{eq:Jerad-indirect-Gamma1} \\
&& \times \frac{\int_0^\infty\mathcal{D}_h(E'_h) n^e_{E_e} n^h_{E'_h} \mathcal{A}_{\sqrt{2} \Gamma }(\Delta E-E_e-E'_h) {\rm d}E'_h}{\iint \mathcal{D}_e(E'_e)\mathcal{D}_h(E'_h) \mathcal{A}_{\sqrt{2} \Gamma}(\Delta E-E'_e-E'_h)  {\rm d}E'_e{\rm d}E'_h} , \nonumber  \\
&& J^{h, {\rm rad}}_{E_h}= \frac{1}{w} \times \frac{c}{2 \pi ^2} \left( \frac{1}{\hbar c}\right)^3 \times  \int_{0}^{\infty}  {\rm d}E  \ E^2  \label{eq:Jhrad-indirect-Gamma1} \\
&& \times \frac{\int_0^\infty\mathcal{D}_e(E'_e) n^e_{E'_e} n^h_{E_h} \mathcal{A}_{\sqrt{2} \Gamma }(\Delta E-E'_e-E_h) {\rm d}E'_e}{\iint \mathcal{D}_e(E'_e)\mathcal{D}_h(E'_h) \mathcal{A}_{\sqrt{2} \Gamma}(\Delta E-E'_e-E'_h)  {\rm d}E'_e{\rm d}E'_h}  , \nonumber
\end{eqnarray}
for the electrons and holes, respectively (where the approximation $1-n^e_{E_e} -n^h_{E_h}  \approx 1$ was used). The above expressions can be simplified further by introducing the dimensionless functions $\Phi(x)$ and $\Theta(x,y)$:
\begin{eqnarray}
\Phi(x) &\equiv& \int_0^\infty \frac{{\rm d} s}{\sqrt{\pi}} s^2 e^{-(s-x)^2} \nonumber \\ 
 &=&\frac{x e^{-x^2}}{2 \sqrt{\pi}}+ \frac{ (1 + 2 x^2) (1 + {\rm erf}(x))}{4}, \\
\Theta (x, y) &\equiv& \int_{-x}^{\infty} \frac{(s+x)^2}{\Phi(s) } \times \frac{e^{-(s-y)^2}}{\sqrt{\pi}}  {\rm d}s ,
\end{eqnarray}
where ${\rm erf}(x)$ is the Gaussian error function and $\Psi(x)$ has following asymptotic forms:
\begin{eqnarray}
\Phi(x \gg 1) =x^2, \quad \Phi(x \ll -1) = \frac{e^{-x^2}}{4 \sqrt{\pi} |x|^3}.
\end{eqnarray}
Using these functions, Eq.~(\ref{eq:Jerad-indirect-Gamma1}) and Eq.~(\ref{eq:Jhrad-indirect-Gamma1}) can be rewritten as:
\begin{eqnarray}
&& J^{e, {\rm rad}}_{E_e}=  \frac{8}{\pi d_e}\left( \frac{c/w}{2 \pi ^2}\right) \left( \frac{1}{\hbar c}\right)^3 n^e_{E_e} \nonumber \\
&&\qquad \times  \int_0^\infty \sqrt{E_h} \Theta \left( \frac{E_g}{\sqrt{2}\Gamma }, \frac{E_e+E_h}{\sqrt{2}\Gamma } \right)  n^h_{E_h} {\rm d}E_h, \quad \label{eq:Je-rad-renew} \\
&& J^{h, {\rm rad}}_{E_h}=  \frac{8}{\pi d_h}\left( \frac{c/w}{2 \pi ^2}\right) \left( \frac{1}{\hbar c}\right)^3 n^h_{E_h} \nonumber \\
&&\qquad \times  \int_0^\infty \sqrt{E_e} \Theta \left( \frac{E_g}{\sqrt{2}\Gamma }, \frac{E_e+E_h}{\sqrt{2}\Gamma } \right)  n^e_{E_e} {\rm d}E_e. \quad \label{eq:Jh-rad-renew}
\end{eqnarray}
Within the limit from $\Gamma \to +0$, we find that $\Theta \to (E_g+E_e+E_h)^2/(E_e+E_h)^2$ in the integrands, which reproduces the original expressions of Eq.~(\ref{eq:Jerad-indirect}) and Eq.~(\ref{eq:Jhrad-indirect}). Equations~(\ref{eq:Je-rad-renew}) and (\ref{eq:Jh-rad-renew}), when derived in this way, are the explicit forms of the radiation loss rates for indirect gap semiconductor absorbers.

A similar analysis is also applied to direct gap semiconductor absorbers.
We find that the expressions in Eq.~(\ref{eq:Jerad-direct}) and Eq.~(\ref{eq:Jhrad-direct}) are modified by the broadening effect as follows:
\begin{eqnarray}
&&J^{e,{\rm rad}}_{E_e}=\frac{c}{2 \pi ^2 w} \left( \frac{1}{\hbar c}\right)^3    \frac{\left( 1+\frac{m_e^\ast}{m_h^\ast}\right)}{d_e \sqrt{E_e}} \label{eq:Jerad-direct-renew}  \\
 && \quad \times 
\int_0 ^\infty E^2 \frac{\sqrt{E_{eh}}  \mathcal{A}_{\sqrt{2}\Gamma}(\Delta E-E_{eh}) n^e_{E_e} n^h_{E_h}}{\int_0 ^\infty \sqrt{E_{eh}'} \mathcal{A}_{\sqrt{2}\Gamma}(\Delta E-E_{eh}') {\rm d} E_{eh}' }  {\rm d}E, \qquad \nonumber \\
&&J^{h,{\rm rad}}_{E_e}=\frac{c}{2 \pi ^2 w} \left( \frac{1}{\hbar c}\right)^3    \frac{\left( 1+\frac{m_h^\ast}{m_e^\ast}\right)}{d_h \sqrt{E_h}} \label{eq:Jhrad-direct-renew}  \\
 && \quad \times 
\int_0 ^\infty  E^2 \frac{\sqrt{E_{eh}}  \mathcal{A}_{\sqrt{2}\Gamma}(\Delta E-E_{eh}) n^e_{E_e} n^h_{E_h}}{\int_0 ^\infty \sqrt{E_{eh}'} \mathcal{A}_{\sqrt{2}\Gamma}(\Delta E-E_{eh}') {\rm d} E_{eh}' } {\rm d}E, \qquad \nonumber 
\end{eqnarray} 
where we used the approximation $1-n^e_{E_e} -n^h_{E_h}  \approx 1$.
Here, $\Delta E \equiv E-E_g$ and $m_e^\ast E_e=m_h^\ast E_h$ for direct gap semiconductors, and $E_{eh} \equiv E_e+E_h$.

{\it For the carrier extraction rate  $J^{e(h),{\rm out}}_{E_{e(h)}}$} --- inclusion of the broadening effect in this term is straightforward. The incoming particle number rates from the electrodes into the absorber are modified because the single particle states have a spectral width. 
After modification, we obtain
\begin{eqnarray}
J^{e,{\rm out}}_{E_{e}}&=&\frac{1}{\tau_{\rm out}} \left( n^e_{E_e}-\tilde{f}^{F}_{\mu_c, \beta_c}(E_g+E_e) \right), \label{eq:Jeout-renew} \\
J^{h,{\rm out}}_{E_{h}}&=&\frac{1}{\tau_{\rm out}} \left( n^h_{E_h}-\tilde{f}^{F}_{-\mu_v, \beta_c}(E_h) \right), \label{eq:Jhout-renew}
\end{eqnarray}  
where we assume that $\tau^{e(h)}_{\rm out}=\tau_{\rm out}$, and the Fermi-Dirac distribution convolved with the lineshape function is given by
\begin{eqnarray}
\tilde{f}^{F}_{\mu, \beta}(E)&=&(f^{F}_{\mu, \beta} \ast \mathcal{A}_\Gamma)(E) \nonumber \\
&=&\int_{-\infty}^\infty f^{F}_{\mu, \beta}(E') \mathcal{A}_\Gamma(E-E')  {\rm d}E'. \label{eq:ConvolvedFDfunc}
\end{eqnarray}

{\it For the phonon scattering rate: $\left.\frac{d}{dt}n^{e(h)}_{E_{e(h)}}\right|_{\rm phonon}$} --- inclusion of the broadening effect replaces the $\delta$-functions in Eq.~(\ref{eq:Je-themal0}) and Eq.~(\ref{eq:Jh-themal0}) with the convolved lineshape functions as follows:
\begin{eqnarray}
 && \delta \bigl(E_{e(h)}(k)-E_{e(h)}(k-q)-E_q^{ph} \bigr) \nonumber \\
&\to&  \mathcal{A}_{\sqrt{2} \Gamma}  \bigl(E_{e(h)}(k)-E_{e(h)}(k-q)-E_q^{ph} \bigr),  \\
 &&   \Bigl( =\int_{-\infty}^{\infty}{\rm d}E_1  \int_{-\infty}^{\infty} {\rm d}E_2  \  \delta \bigl(E_1-E_2-E_q^{ph} \bigr) \nonumber \\
&& \quad \times \mathcal{A}_\Gamma \bigl( E_1-E_{e(h)}(k) \bigr) \mathcal{A}_\Gamma \bigl( E_2-E_{e(h)}(k-q) \bigr)  \Bigr) \nonumber \\
 && \delta \bigl(E_{e(h)}(k+q)-E_{e(h)}(k)-E_q^{ph} \bigr) \nonumber \\
&\to&  \mathcal{A}_{\sqrt{2} \Gamma}  \bigl(E_{e(h)}(k+q)-E_{e(h)}(k)-E_q^{ph} \bigr).
\end{eqnarray}
After the replacement, we can transform the $q$-summation into an integration over polar coordinates, and the angular integration can finally be performed, as was done in Sec.~\ref{Jph}.
The final result is
\begin{eqnarray}
&&\left.\frac{d}{dt}n^{e(h)}_{E_{e(h)}}\right|_{\rm phonon}  \nonumber \\
&=& 
 \frac{-C_{ph}^{e(h)} }{\sqrt{E_{e(h)}}}
 \int_0^{\epsilon^{ph}_{\rm cut}} \epsilon^2{\rm d}\epsilon \int^{E_+}_{E_-} {\rm d}E'
\nonumber \\
& & \mathcal{A}_{\sqrt{2}\Gamma}(E_{e(h)}-E'-\epsilon )   \biggl(    n^{e(h)}_{E_{e(h)}}(1-n_{E'}^{e(h)}) 
 \nonumber \\
&& \times \bigl( 1+f^B_{0, \beta_{\rm ph}} ({\epsilon})\bigr) -n_{E'}^{e(h)}  (1-n^{e(h)}_{E_{e(h)}})  f^B_{0, \beta_{\rm ph}} ({\epsilon}) \biggr)  \nonumber \\
&+&
 \frac{C_{ph}^{e(h)} }{\sqrt{E_{e(h)}}}
 \int_0^{\epsilon^{ph}_{\rm cut}} \epsilon^2{\rm d}\epsilon \int^{E_+}_{E_-} {\rm d}E'
   \label{eq:Jeh-themal-renew}  \\
& & \mathcal{A}_{\sqrt{2}\Gamma}(E_{e(h)}-E' + \epsilon )  \biggl(    n^{e(h)}_{E'}(1-n_{E_{e(h)}}^{e(h)}) 
 \nonumber \\
&& \times  \bigl( 1+f^B_{0, \beta_{\rm ph}} ({\epsilon})\bigr)  -n_{E_{e(h)}}^{e(h)} (1-n^{e(h)}_{E'})  f^B_{0, \beta_{\rm ph}} ({\epsilon}) \biggr)  ,\nonumber
\end{eqnarray}
where we defined $E_{\pm}$ as 
\begin{eqnarray}
E_{\pm} \equiv \left( \sqrt{E_{e(h)}} \pm \epsilon \sqrt{\frac{1}{2 m^\ast_{e(h)}v_A^2}} \right)^2 \ge 0. \label{eq:def-Epm}
\end{eqnarray}
Within the limit from $\Gamma \to +0$ in Eq.~(\ref{eq:Jeh-themal-renew}), the Gaussian function $\mathcal{A}_{\sqrt{2}\Gamma}$ becomes $\delta$ functions. 
The terms that remain after integration of $E'$ should come from the poles $E'=E_{e(h)}\pm \epsilon$, which are located in the interval $E_-<E'(=E_{e(h)}\pm \epsilon)<E_+$. In this way, Eq.~(\ref{eq:Jeh-themal-renew}) safely reproduces Eq.~(\ref{eq:Jeh-themal1}) when $\Gamma = +0$.
Here, we assumed again that $E_{e(h)}> E^{\rm PB}_{e(h)}$, i.e., that the phonon bottleneck effect was neglected as it was in Eq.~(\ref{eq:Jeh-themal1}).

{\it The reason for adoption of the Gaussian lineshape}---
As mentioned earlier in this subsection, we have used a Gaussian lineshape for the microscopic states, despite the fact that it cannot be derived naturally for an ideal model without statistical fluctuations. This lineshape was adopeted for the following technical reason.
We have derived rate equations for the microscopic distribution functions of the carriers in both direct and indirect gap semiconductor absorbers. In the derivation for the indirect gap semiconductors, we introduced a probability density of $p^{e(h)}(E_{e(h)})$ in Eq.~(\ref{eq:pe}) and used it to obtain the generation rate $J^{e(h),{\rm sun}}_{E_{e(h)}}$ and the recombination loss rate $J^{e(h),{\rm rad}}_{E_{e(h)}}$. 
The factor $\int_0^{\Delta E} \mathcal{D}_e(E')  \mathcal{D}_h(\Delta E-E'){\rm d}E'$ that is found in the denominator in each of Eq.~(\ref{eq:Je-sun2}), Eq.~(\ref{eq:NeNh}), and Eq.~(\ref{eq:1-Ne-Nh}) can all be traced back to the same normalization factor in the definition of $p^{e(h)}(E_{e(h)})$ in Eq.~(\ref{eq:pe}), i.e., the number of possible combinations of all electron-hole pairs with $E_e$ and $E_h$ that can emit a photon of energy $E$. 
If the spectral broadening of the microscopic states is included, the normalization factor should then be modified by replacing $\mathcal{D}_{e(h)}(E)$ with $\tilde{\mathcal{D}}_{e(h)}(E)$ (as defined in Eq.~(\ref{eq:Deh-renew})), as shown in Eq.~(\ref{eq:Normalization-renew}).
No problems arise with the mathematics here if Gaussian lineshape function $\mathcal{A}_{\Gamma}(x)$ is used for the lineshape.
However, if we use the Lorentzian lineshape function, 
\begin{eqnarray}
\mathcal{A}^{L}_{\Gamma}(x) \equiv \frac{1}{\pi}\frac{\Gamma}{x^2+\Gamma^2},
\end{eqnarray}
the situation changes. Using a similar analysis, the normalization factor is calculated to be
\begin{eqnarray}
&& \int_{-\infty}^{\infty}  \tilde{\mathcal{D}}_e(E')\tilde{\mathcal{D}}_h(\Delta E-E') {\rm d}E' \nonumber \\
&=& \int_{-\infty}^\infty {\rm d}E' \int_{0}^\infty {\rm d}E_e \int_{0}^\infty {\rm d}E_h  \mathcal{D}_e(E_e) \mathcal{A}^L_\Gamma (E'-E_e) \nonumber \\
&& \times  \mathcal{D}_h(E_h) \mathcal{A}^L_\Gamma (\Delta E-E'-E_h) \nonumber \\
&=& \int_{0}^\infty {\rm d}E_e \int_{0}^\infty {\rm d}E_h  \mathcal{D}_e(E_e)  \mathcal{D}_h(E_h) \nonumber \\
&& \qquad \qquad  \times  \mathcal{A}^L_{2\Gamma} (\Delta E-E_e-E_h) .  \label{eq:NormalizationL-renew}
\end{eqnarray}
In the last equation, we used a general property of the convolution of Lorentzian functions:
\begin{eqnarray}
\mathcal{A}^L_{\Gamma_1}\ast \mathcal{A}^L_{\Gamma_2}(E)&\equiv&
 \int_{-\infty}^\infty {\rm d}E'\mathcal{A}^L_{\Gamma_1}(E') \mathcal{A}^L_{\Gamma_2}(E-E') \nonumber \\
&=& \mathcal{A}^L_{\Gamma_1+\Gamma_2}(E).
\end{eqnarray}
The integration can also be performed by changing the integration variable to $E_{eh}\equiv E_e+E_h$ and $E_h$, which results in the following divergence:
\begin{eqnarray}
&&\int_{-\infty}^{\infty}  \tilde{\mathcal{D}}_e(E')\tilde{\mathcal{D}}_h(\Delta E-E') {\rm d}E' \nonumber \\
&\propto& \int_{0}^\infty {\rm d}E_{eh} \int_{0}^{E_{eh}} {\rm d}E_h  \sqrt{E_h(E_{eh}-E_h)}  \mathcal{A}^L_{2\Gamma} (\Delta E-E_{eh})  \nonumber \\
&=&\frac{\pi}{8}\int_{0}^\infty E_{eh}^2  \mathcal{A}^L_{2\Gamma} (\Delta E-E_{eh})   {\rm d}E_{eh} \nonumber \\
&\propto& \int_{0}^\infty \frac{E_{eh}^2}{E_{eh}^2+(2\Gamma)^2} {\rm d}E_{eh}  = +\infty. \label{eq:UVdiv}
\end{eqnarray} 
This divergence represents a clear artifact that is derives from the simplification of our model of the semiconductor carriers.
We adopted an effective two-band model with infinite bandwidths for carriers. Because of these infinite bandwidths, there is no upper domain bound for integration over $E_{eh}$ in Eq.~(\ref{eq:UVdiv}).
This prevents us from defining the probability measure for $p^{e(h)}(E_{e(h)})$ and our formulation thus fails when we use the Lorentzian lineshape function and the infinite bandwidths for the carriers. 
Of course, this problem can be avoided by introducing an effective bandwidth parameter that should exist naturally.
However, the introduction of an additional bandwidth parameter can then reduce the benefits of our simple two-band model with the infinite bandwidths. 
The final results may also depend on the bandwidth parameter, the definition of which remains unclear.
For these reasons, we have used the Gaussian lineshape for the spectral function, not having encountered such an artifact, and have thus benefitted greatly from use of our simplified model.
In realistic devices, the semiconductor absorbers are not ideally prepared with structural imperfections to cause statistical fluctuations. As a result, their spectral lineshapes will be more or less Gaussian-like, at least in the tails (which could also be Voigt functions).
Even when the Lorentzian lineshape is used in the infinite bandwidth model, this divergence problem does not occur for direct gap semiconductors. 
Despite the above discussion on possible changes in the formulation, we still expect that the selection of the lineshape functions will not strongly affect the result or the main conclusion.

\subsection{\label{efficiency} Macroscopic properties: current, conversion efficiency, and energy flow}
\begin{figure}[!t]
\centering
\includegraphics[width=0.5\textwidth]{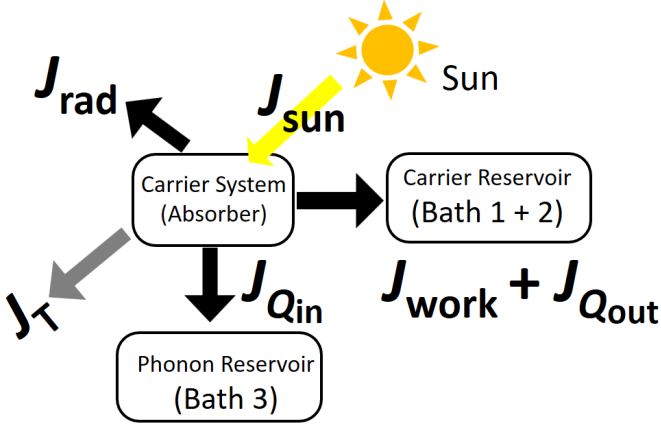}
\caption{Energy current from the Sun $J_{\rm sun}$, which is shared by five energy current channels: radiation $J_{\rm rad}$, light transmission $J_{\rm T}$, output work $J_{\rm work}$, heat passing into the electrodes $J_{Q_{\rm out}}$, and heat passing into the absorber crystal lattice $J_{Q_{\rm in}}$.}  
\label{fig7}  
\end{figure}

The rate equations used to determine the microscopic carrier distribution functions were derived in the preceding subsections (from Sec.~\ref{Jsun} to Sec.~\ref{broadening}).
In this section, we briefly summarize the method used to calculate macroscopic quantities such as the output charge current $I$, the conversion efficiency $\eta$, and the energy flows into the different channels, denoted by $J_{X}$ ($X=$sun, T, rad, work, $Q_{\rm in}$, and $Q_{\rm out}$ are defined below and are also shown in Fig.~\ref{fig7}) in terms of the distribution functions that are to be obtained.  

The charge current $I$ is defined as
\begin{eqnarray}
I&=&|e| \sum_{k} J_{E_e(k)}^{e, {\rm out}} 
= |e| \sum_{k} J_{E_h(k)}^{h, {\rm out}}, \nonumber \\
&=& |e|  \int_0^\infty \mathcal{V}  \mathcal{D}_e(E_e) J_{E_e}^{e, {\rm out}}   {\rm d}E_{e}, \label{eq:Ecurrent}
\end{eqnarray} 
which represents the total charge per unit time (where $|e|$ is the elementary charge) output by an absorber of volume $\mathcal{V} (=\mathcal{A}w)$ to an electrode.
Here $J_{E_{e}}^{{e}, {\rm out}} $ and $J_{E_{h}}^{{h}, {\rm out}} $ are the functions of the distribution functions given by Eq.~(\ref{eq:Jeout1}) and Eq.~(\ref{eq:Jhout1}) without the broadening effect, and of the functions given by Eq.~(\ref{eq:Jeout-renew}) and Eq.~(\ref{eq:Jhout-renew}) with the broadening effect, respectively.
The second equation in Eq.~(\ref{eq:Ecurrent}) comes from the steady-state condition based on the total number of charges in the absorber (and will be mentioned again as Eq.~(\ref{eq:charge-neutrality2}) in Sec.~\ref{SecIII}).
Given that the conduction electrons are extracted to the electrode with chemical potential $\mu_c$ and the valence holes are extracted to the electrode with potential $\mu_v$ ($-\mu_v$ for holes), the total output energy per unit time is given by
\begin{eqnarray}
\mathcal{P}_{\rm work}=(\mu_c-\mu_v) I=VI,
\end{eqnarray}
which defines the conversion efficiency as 
\begin{eqnarray}
\eta (\%) =\frac{\mathcal{P}_{\rm work}}{\mathcal{A} \times \int_{0}^{\infty} E  j^{\rm sun}(E) {\rm d}E  } \times 100.
\end{eqnarray}
This expression indicates that the conversion efficiency is dependent on the absorber thickness $w$ because $I$ and $\mathcal{P}_{\rm work}$ are both proportional to $\mathcal{V}$ from Eq.~(\ref{eq:Ecurrent}).

The microscopic carrier distribution function offers more detailed information about the energy balance of the solar cells.
This information is very helpful in understanding what occurs inside solar cells during the energy conversion processes, which could also be addressed experimentally ~\cite{Chen}. 
The energy current $J_{X}$ (per unit time per unit area) that flows into each channel $X$ can be evaluated separately using the steady-state solution as follows (see also Fig.~\ref{fig7}).

The energy current per unit time per unit area that is carried by the photons that illuminate the solar cell is 
\begin{eqnarray}
J_{\rm sun}= \int_{0}^{\infty} E  j^{\rm sun}(E) {\rm d}E.
\end{eqnarray}
The energy flow from sunlight is shared by different flow channels (five channels are present in this model), where
\begin{eqnarray}
J_{\rm sun}=J_{\rm T}+J_{\rm rad}+J_{\rm work}+J_{Q_{\rm out}}+J_{Q_{\rm in}}. \label{eq:energy balance}
\end{eqnarray} 

The proportion of the solar energy, $J_{\rm T}$, per unit time per unit area, that is transmitted is given by 
\begin{eqnarray}
J_{\rm T}= \int^{E_g}_{0} E  j^{\rm sun}(E) {\rm d}E.
\end{eqnarray}
Here, the absorption spectrum of the absorber is approximated simply using the step function $\alpha(E)=\alpha_0 \theta(E-E_g)$ under the assumption of perfect absorption $\alpha_0 w \gg 1$ (or a large effective light path $l_{\rm eff} \gg 1/\alpha_0$ with light trapping textures~\cite{Yablonovitch}). 

The energy that is radiated outside the solar cell per unit time per unit area is then given by
\begin{eqnarray}
J_{\rm rad}&=& \frac{ c}{2 \pi ^2} \left( \frac{1}{\hbar c}\right)^3 \int_{0}^{\infty}
E^3  \langle n^e n^h \rangle_{\Delta E} {\rm d}E,
\end{eqnarray}
where $\Delta E \equiv E-E_g$, and $1-n^e_{E_e} - n^h_{E_h}\approx 1$ is used.
When the broadening effect is taken into accoint, $\langle n^e n^h \rangle_{\Delta E}$ in the integrand is given by
\begin{eqnarray}
&&\langle n^e n^h \rangle_{\Delta E}=\int_0^\infty \int_0^\infty p_{E_e, E_h}^{\Delta E} n^e_{E_e} n^h_{E_h} {\rm d}E_e{\rm d} E_h , \qquad \label{eq:NeNh-renew-indirect} 
\end{eqnarray}
where $p_{E_e, E_h}^{\Delta E} $ is a weight function that is defined by
\begin{eqnarray}
&&p_{E_e, E_h}^{\Delta E}  \\
&\equiv &
\frac{\mathcal{D}_e(E_e)\mathcal{D}_h(E_h)  \mathcal{A}_{\sqrt{2} \Gamma }(\Delta E-E_e-E_h)  }
{\iint \mathcal{D}_e(E_e)\mathcal{D}_h(E_h) \mathcal{A}_{\sqrt{2} \Gamma}(\Delta E-E_e-E_h)  {\rm d}E_e{\rm d}E_h} \nonumber
\end{eqnarray}
for indirect gap semiconductors.
For the direct gap semiconductors, $\langle n^e n^h \rangle_{\Delta E}$ is given by
\begin{eqnarray}
&&\langle n^e n^h \rangle_{\Delta E}= \int_0^\infty p_{E_{eh}}^{\Delta E} n^e_{E_e} n^h_{E_h} {\rm d}E_{eh} , \qquad \label{eq:NeNh-renew-direct} 
\end{eqnarray}
with a weight function that is defined by
\begin{eqnarray}
p_{E_{eh}}^{\Delta E} \equiv
\frac{\sqrt{E_{eh}}  \mathcal{A}_{\sqrt{2} \Gamma }(\Delta E-E_{eh})  }{\int_0^\infty \sqrt{E'_{eh}} \mathcal{A}_{\sqrt{2} \Gamma}(\Delta E-E'_{eh})  {\rm d}E'_{eh} } ,
\end{eqnarray}
where $\Delta E \equiv E-E_g$, $m_e^\ast E_e=m_h^\ast E_h$, and $E_{eh} \equiv E_e+E_h$.
In a similar manner to the discussion in Sec.~\ref{broadening}, the expressions above can be used in the limit where $\Gamma =+0$ by changing the lineshape functions in the integrand into delta functions.

$J_{\rm work}$ is the energy that is extracted as work from the solar cell (per unit time per unit area) and transferred via the charge current, which we already evaluated earlier, and is given by
\begin{eqnarray}
J_{\rm work}&=& P_{\rm work}/\mathcal{A} \nonumber \\
&=&w|e|V \int_0^\infty   \mathcal{D}_e(E_e) J_{E_e}^{e, {\rm out}}   {\rm d}E_{e}.
\end{eqnarray}

$J_{Q_{\rm out}}$ is the heat that is carried by the charge current and lost in the electrodes outside the cell per unit time per unit area.
It can be calculated as:
\begin{eqnarray}
&& J_{Q_{\rm out}} \nonumber \\
&=&\frac{w}{\tau_{\rm out}}\iint  \mathcal{D}_{E_e}^e E \mathcal{A}_{\Gamma}(\Delta E-E_e)\left( n^{e}_{E_e}-f^{F}_{\mu_c, \beta_{c}}(E) \right)  {\rm d}E_e{\rm d}E  \nonumber \\
&+& \frac{w}{\tau_{\rm out}}\iint \mathcal{D}_{E_h}^h E \mathcal{A}_{\Gamma}(E-E_h)\left( n^{h}_{E_h}-f^{F}_{-\mu_v, \beta_{c}}(E) \right) {\rm d}E_h{\rm d}E  \nonumber \\
&-& J_{\rm work}, \label{eq:JQout}
\end{eqnarray} 
where $\Delta E \equiv E-E_g$, and the integration runs over the ranges $-\infty<E<\infty$ and $0<E_{e(h)} <\infty$.

The remaining channel for the energy loss in our model is the absorber thermalization loss that is lost in the phonon reservoir (Bath 3 in Fig.~\ref{fig1}).
$J_{Q_{\rm in}}$ is this heat current, which is lost in the phonon bath, inside the cell per unit time per unit area.
It is convenient to determine $J_{Q_{\rm in}}$ using the energy conservation law given in Eq.~(\ref{eq:energy balance}),
\begin{eqnarray}
J_{Q_{\rm in}}=J_{\rm sun}-J_{\rm T}-J_{\rm rad}-J_{\rm work}-J_{Q_{\rm out}}, \label{eq:thermalization loss}
\end{eqnarray}
because all other terms on the right-hand side have been given above.
It can also be evaluated directly using the phonon scattering rate in the rate equation; however, this complicates the calculation. 

\section{\label{SecIII} Basic properties and classification of the parameter regime}
In the previous section, the rate equations that determine the microscopic distribution functions were derived based on our nonequilibrium model.
In this section, before we describe the numerical simulation, we give the basic properties related to the particle number conservation laws that must be preserved in any steady states for the equations.
This information can then be used in the first screening to validate the numerical results obtained.
In the latter part of this section, we will determine the parameter regime in which the solar cells will work under or out of the detailed balance condition.
The parameter regime can be classified from the equation itself without fully solving for it. 

\subsection{\label{intraband scattering} Conservation of the total number of carriers within the band during phonon scattering processes}
The rate equations in Eq.~(\ref{eq:e-balance}) and Eq.~(\ref{eq:h-balance}) represent the net rates for the numbers of particles coming into the microscopic states with energies $E_e$ and $E_h$.
Therefore, as shown in the previous section (Sec.~\ref{efficiency}), the summation of ${\rm d} n^{e(h)}_{E_{e(h)}} /{\rm d}t$ over all microscopic states gives the total net number of electrons (or holes) coming into the absorber (or the cell) per unit time.
In the steady state, the net total rate vanishes as a result of the balance between the four terms that are related to generation by sunlight absorption, the radiative recombination loss, extraction to the electrodes, and phonon scattering, on the right-hand side of the rate equation.
Among these four terms, the last one, which leads to intraband carrier thermalization, should not change the total number of carriers contained within the band.
This basic property must be preserved in our model formulation, which can be checked using the general expression given in Eq.~(\ref{eq:Jeh-themal-renew}) (with the broadening effect) as shown below.

The net change in the total number of electrons (holes) in the absorber due to the phonon scattering processes per unit time is given by
\begin{eqnarray}
\mathcal{V} \int_0^\infty \mathcal{D}_{e(h)}(E_{e(h)}) \left.\frac{d}{dt}n^{e(h)}_{E_{e(h)}}\right|_{\rm phonon}   {\rm d}E_{e(h)}. \label{eq:IntrabandScatteringRate}
\end{eqnarray} 
By inserting Eq.~(\ref{eq:Jeh-themal-renew}) into Eq.~(\ref{eq:IntrabandScatteringRate}), we find that the integrand of the $\epsilon$-integration is proportional to
\begin{eqnarray}
\int_0^\infty  {\rm d}E_{e(h)} \int^{E_+}_{E_-} {\rm d}E' \left( I^\epsilon_{E_{e(h)},E'} -I^\epsilon_{E',E_{e(h)}} \right) , \label{eq:I12-I21}
\end{eqnarray}
where
\begin{eqnarray}
&& I^\epsilon_{E_{e(h)},E'}  \\
&\equiv& \mathcal{A}_{\sqrt{2}\Gamma}(E_{e(h)}-E'-\epsilon )   \biggl(    n^{e(h)}_{E_{e(h)}}(1-n_{E'}^{e(h)}) 
 \nonumber \\
&& \times \bigl( 1+f^B_{0, \beta_{\rm ph}} ({\epsilon})\bigr) -n_{E'}^{e(h)}  (1-n^{e(h)}_{E_{e(h)}})  f^B_{0, \beta_{\rm ph}} ({\epsilon}) \biggr) . \nonumber 
\end{eqnarray}
From the definition of $E_{\pm}$ given in Eq.~(\ref{eq:def-Epm}), the integration domain, $E_-<E'<E_+$, is equivalent to
\begin{eqnarray}
&& E_{e(h)}<  \left( \sqrt{E'}+ \epsilon \sqrt{\frac{1}{2 m^\ast_{e(h)}v_A^2}} \right)^2, \\
&& E_{e(h)}>  \left( \sqrt{E'}- \epsilon \sqrt{\frac{1}{2 m^\ast_{e(h)}v_A^2}} \right)^2,
\end{eqnarray}
which are both symmetrical with respect to interchange of the following variables: $E' \leftrightarrow E_{e(h)}$.
In this way, we find that the two contributions in Eq.~(\ref{eq:I12-I21}) cancel perfectly after integration.
This ensures that the total number of carriers is preserved within the band during the carrier thermalization process.

\subsection{\label{charge neutrality} Conservation of total charge and charge neutrality}
The rate equation for the net total charge in the absorber, $Q_{\rm tot} (\equiv |e|\sum_k (n^h_{E_h (k)}-n^e_{E_e (k)}))$, can also be obtained using the microscopic rate equations. 
Summation over all states and the difference between those states for the electrons and holes give   
\begin{eqnarray}
&&\frac{d}{dt}Q_{\rm tot}  \label{eq:charge-neutrality} \\
&=&-|e|\mathcal{V}\int_0^\infty \mathcal{D}_e(E_e)(J^{e,{\rm sun}}_{E_e}-J^{e,{\rm rad}}_{E_e}-J^{e,{\rm out}}_{E_e}) {\rm d}E_e \nonumber \\
&&+|e|\mathcal{V}\int_0^\infty \mathcal{D}_h(E_h)(J^{h,{\rm sun}}_{E_h}-J^{h,{\rm rad}}_{E_h}-J^{h,{\rm out}}_{E_h}) {\rm d}E_h,  \nonumber 
\end{eqnarray}
where the phonon scattering terms are absent, in line with the discussion in Sec.~\ref{intraband scattering}.
Additionally, the terms for photon absorption and recombination radiation on the right-hand side should all cancel. 
This should be true because each single photon absorption and emission process generates or loses one electron-hole pair with no changes in the total charge.  
This statement can also be verified easily in our formulation using the general expressions for the generation rate (Eq.~(\ref{eq:Je-sun-direct}) and Eq.~(\ref{eq:Jh-sun-direct}) for direct gap semiconductors, and Eq.~(\ref{eq:Je-sun2}) and Eq.~(\ref{eq:Jh-sun2}) for indirect gap semiconductors) and the radiative recombination rate (Eq.~(\ref{eq:Jerad-direct-renew}) and Eq.~(\ref{eq:Jhrad-direct-renew}) for direct gap semiconductors, and Eq.~(\ref{eq:Je-rad-renew}) and Eq.~(\ref{eq:Jh-rad-renew}) for indirect gap semiconductors). As a result, Eq.~(\ref{eq:charge-neutrality}) can be rewritten as
\begin{eqnarray}
\frac{d}{dt}Q_{\rm tot}&=&|e|\mathcal{V} \Bigl(\int_0^\infty \mathcal{D}_e(E_e) J^{e,{\rm out}}_{E_e} {\rm d}E_e  \nonumber \\
 && -\int_0^\infty \mathcal{D}_h(E_h)J^{h,{\rm out}}_{E_h} {\rm d}E_h \Bigr)=0,   \label{eq:charge-neutrality2} 
\end{eqnarray}
where the steady state has been assumed in the second equation.
 The above equation, which shows that the net extraction rates for the electrons and holes are balanced in the steady state, was used in Eq.~(\ref{eq:Ecurrent}) in the previous section.
Additionally, when using Eq.~(\ref{eq:Jeout-renew}) and Eq.~(\ref{eq:Jhout-renew}) along with the definition of the total charge, the total net charge that is present in the absorber is determined from Eq.~(\ref{eq:charge-neutrality2}) under steady-state conditions and is given by 
\begin{eqnarray}
\frac{Q_{\rm tot}}{\tau_{\rm out}}&=&\frac{|e|\mathcal{V}}{\tau_{\rm out}} \Bigl(
\int_0^\infty \mathcal{D}_h(E_h) \tilde{f}^{F}_{-\mu_v, \beta_c}(E_h) {\rm d}E_h \nonumber \\
&& -\int_0^\infty \mathcal{D}_e(E_e) \tilde{f}^{F}_{\mu_c-E_g, \beta_c}(E_e) {\rm d}E_e
\Bigr). \label{eq:TotalCharge}
\end{eqnarray}
From this result, we see that the total charge that is present in the absorber is solely determined by the cell distribution functions (apart from the broadening effect).
In the numerical simulations presented here, we have focused only on the special case of charge neutrality, where $Q_{\rm tot}=0$. 
When the condition $n^{e(h)}_{E_{e(h)}}\ll 1$ is satisfied at low carrier densities, the charge neutrality condition then relates the chemical potentials in the electrodes, $\mu_c$ and $\mu_v$, as follows:
\begin{eqnarray}
\frac{\mu_c+\mu_v}{2}=\frac{E_g}{2} +\frac{\beta_c^{-1}}{2} \ln{\frac{d_h}{d_e}}. \label{eq:neutrality condition1}
\end{eqnarray}
Therefore, using the voltage between the electrodes ($|e|V \equiv \mu_c-\mu_v$), the chemical potentials can be given by 
\begin{eqnarray}
\mu_c&=&\left( E_g+\beta_c^{-1} \ln (d_h/d_e)+|e|V \right)/2,  \label{eq:neutrality condition2}\\
\mu_v &=& \left( E_g+\beta_c^{-1} \ln (d_h/d_e)-|e|V \right)/2, \label{eq:neutrality condition3}
\end{eqnarray}
to enable charge neutrality to be realized in the absorber.

\subsection{\label{classification parameter} Classification of the parameter regime}
In Sec.~\ref{SecII}, we saw that the rate equation has two characteristic times: $\tau_{\rm out}$ for carrier extraction (from Eq.~(\ref{eq:tout-e}) and Eq.~(\ref{eq:tout-h})) and $\tau_{\rm ph}^{e(h)}$ for carrier thermalization (from Eq.~(\ref{eq:PhononScatteringRate})), which will be determinate for the solar cell properties.
As noted in Sec.~\ref{SecI}, the SQ theory assumes that these two characteristic times should be fast enough for the detailed balance analysis to be applicable, whereas the quantitative issue of how short these times should be remains to be solved.
Here, using the rate equations that were derived using the nonequilibrium approach, we address this issue.

\begin{figure}[!t]
\centering
\includegraphics[width=0.5\textwidth]{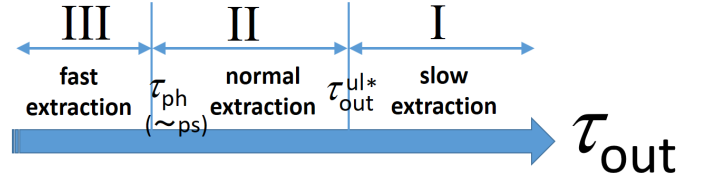}
\caption{Classification of parameter space $0<\tau_{\rm out}<+\infty$ into three different regimes: (I) slow extraction, where $\tau_{\rm out}> \tau_{\rm out}^{{\rm ul}\ast}$; (II) normal extraction, where $\tau_{\rm ph}^{e(h)}<\tau_{\rm out}< \tau_{\rm out}^{{\rm ul}\ast}$; and (III) fast extraction, where $\tau_{\rm out}<\tau_{\rm ph}^{e(h)}$}  
\label{fig8} 
\end{figure}

In the following discussion, we consider $\tau_{\rm out}$ to be a free parameter, while $\tau_{\rm ph}^{e(h)}$ is the given material parameter and is roughly of picosecond order (although the parameter may be modified to a certain degree by introduction of phononic nanostructures ~\cite{Yu, Hopkins}).
We therefore focus on the parameter space given by $\tau_{\rm out}$ (in a similar manner to parameterization using the conductance in hot carrier solar cells~\cite{Takeda2}). 
As shown in Fig.~\ref{fig8}, we divide the parameter space given by $0<\tau_{\rm out}<+\infty$ into three different regimes. 

In regime III, i.e., the fast extraction regime where $\tau_{\rm out} < \tau_{\rm ph}^{e(h)}$, solar cell operation is likely to be different from that of conventional solar cells because carrier extraction before thermalization with the lattice will become possible~\cite{Nozik, Wurfel2}. 
In real devices, fast extraction within the ps scale requires an ultrathin absorber. With an average estimated velocity at room temperature of $v_e = \sqrt{2 m_e^\ast \times 3 k_B T_c/2} \sim 10^5$m/s, the carriers can travel a maximum of 0.1 $\mu$m within 1 ps. 
Because the extraction time cannot be less than the travel time from the center of the absorber (where photogeneration occurs) to its surface (where extraction occurs), $\tau_{\rm out}$ of less than 1 ps requires absorbers with less than sub-$\mu$m thickness in planar solar cells.
The situation can hardly be realized in Si solar cells with low absorption coefficients (a thickness of 10 $\mu$m will be required even with light trapping for perfect absorption which was assumed here). To achieve fast extraction in Si solar cells, a meander-like structure could be used (see Sec.~7.4 in \cite{Wurfel text}). 
However, for direct gap semiconductors with higher absorption coefficients, sub-$\mu$m-scale absorbers could possibly be used to provide efficient solar cells. 
This is why hot carrier solar cells should use ultrathin nanostructures with strongly absorbing materials, such as GaAs, as described in the next section.
While we acknowledge these realistic issues, we will however proceed with further discussions and simulation of fast extraction regime III, given that this scenario could even be realized in Si solar cells, to highlight the general properties of solar cells operating in this fast extraction regime. 

The SQ theory assumes fast carrier extraction, where the condition $\tau_{\rm out} < \tau_{\rm ph}^{e(h)}$ is not required.
An upper limit on the extraction time, $\tau_{\rm out}^{{\rm ul} \ast} (>\tau_{\rm ph}^{e(h)})$, should therefore exist, above which the SQ theory will fail to predict the conversion efficiency limit.
We can therefore define these two regimes separately using a boundary as the normal extraction regime (II) and the slow extraction regime (I).
 In regime II, the SQ theory can be used to predict the solar cell properties.
The boundary time $\tau_{\rm out}^{\rm ul \ast}$ between regimes I and II, which will be dependent on device parameters such as the material and thickness of the absorber, can be evaluated as shown in the remainder of this section.

In regimes I and II, we can safely assume that the thermalization (intraband carrier cooling) is completed within the absorber because the phonon scattering rate dominates the other terms in the rate equations, Eq.~(\ref{eq:e-balance}) and Eq.~(\ref{eq:h-balance}).
In this case, the carrier distribution function in the cell is given by the Fermi-Dirac distribution function in Eq.~(\ref{eq:FDfuncPartialEquilibrium}), while the chemical potentials in the cell can differ from those in the electrodes, i.e., $\mu^{\rm cell}_{e(h)} \ne \mu_{e(h)}$. 
When the function in Eq.~(\ref{eq:FDfuncPartialEquilibrium}) has been assumed, we can insert $\left.\frac{d}{dt}n^{e(h)}_{E_{e(h)}}\right|_{\rm phonon}=0$ into the rate equations in Eq.~(\ref{eq:e-balance}) and Eq.~(\ref{eq:h-balance}). 
This is what we observed earlier in Sec.~\ref{broadening}.

We should also reconsider the meaning of the relation $\mu^{\rm cell}_{e(h)} \ne \mu_{e(h)}$ here.
This means that the carrier distribution functions in the cell differ from those in the electrodes, while they are assumed to be equal in the SQ theory when calculating the radiative recombination loss. 
Using the difference $\Delta n^{e(h)}_{E_{e(h)}} \bigl( \equiv n^{e(h)}_{E_{e(h)}}-f^{F}_{\mu_{e(h)}, \beta_c}(E_{e(h)}) \bigr)$, the microscopic rate equations for the steady states can be rewritten as
\begin{eqnarray}
J^{e(h),{\rm out}}_{E_{e(h)}}=\frac{\Delta n^{e(h)}_{E_{e(h)}}}{\tau_{\rm out}}=J^{e(h),{\rm sun}}_{E_{e(h)}}-J^{e(h),{\rm rad}}_{E_{e(h)}}.
\end{eqnarray}
Therefore, the difference is given by
\begin{eqnarray}
\Delta n^{e(h)}_{E_{e(h)}}&=&\tau_{\rm out}\left( J^{e(h),{\rm sun}}_{E_{e(h)}}-J^{e(h),{\rm rad}}_{E_{e(h)}} \right) \nonumber \\
&\approx & \tau_{\rm out} J^{e(h),{\rm sun}}_{E_{e(h)}},
\end{eqnarray}
Where, in the second line, we have used the approximation $J^{e(h),{\rm sun}}_{E_{e(h)}} \gg J^{e(h),{\rm rad}}_{E_{e(h)}}$, which is usually fulfilled at the maximum operating power point of $V=V_{\rm mp}$ (0.779 V for Si and 1.052 V for GaAs at 1 sun when using our 6000 K blackbody Sun), being focused from this point in this subsection.
Now we will determine the condition by which the SQ conversion efficiency limit can be modified.
When $\tau_{\rm out}$ increases inside the regime II and gradually approaches I, we can fix the maximum operating power point because the maximum power condition given in the SQ theory will at least remain unchanged inside regime II.
The SQ calculation using $n^{e(h)}_{E_{e(h)}} = f^{F}_{\mu_{e(h)}, \beta_c}(E_{e(h)})$ will be justified as long as $\Delta n^{e(h)}_{E_{e(h)}} \ll  f^{F}_{\mu_{e(h)}, \beta_c}(E_{e(h)})$ is satisfied at $V=V_{\rm mp}$, which can be read as
\begin{eqnarray}
\tau_{\rm out} \ll \frac{ f^{F}_{\mu_{e(h)}, \beta_c}(E_{e(h)}) }{ J^{e(h),{\rm sun}}_{E_{e(h)}} } \equiv \tau_{\rm out}^{\rm ul}(E_{e(h)}). \label{eq:tau_out-UL1}
\end{eqnarray}
When $\tau_{\rm out}> \tau_{\rm out}^{\rm ul}(E_{e(h)})$, the conversion efficiency limit can then differ from the limit from the SQ theory.
A sufficient condition for the SQ calculation to be justified is that where Eq.~(\ref{eq:tau_out-UL1}) is fulfilled over an energy range in which there is a non-negligible carrier distribution that makes a relevant contribution to the radiative recombination; this is given approximately by $0<E_{e(h)}<k_B T_c$. 
(Given that $f^{F}_{\mu_{e(h)}, \beta_c}(E_{e(h)})$ decreases exponentially with $E_{e(h)}$, it is almost impossible to satisfy this condition at higher energies.)
In this way, the upper boundary of parameter regime II can be evaluated in principle using $\tau_{\rm out}^{\rm ul}(E_{e(h)})$ in Eq.~(\ref{eq:tau_out-UL1}), e.g. by selecting $E_{e(h)}=k_B T_c$ as a typical carrier energy scale.

\begin{figure}[!t]
\centering
\includegraphics[width=0.45\textwidth]{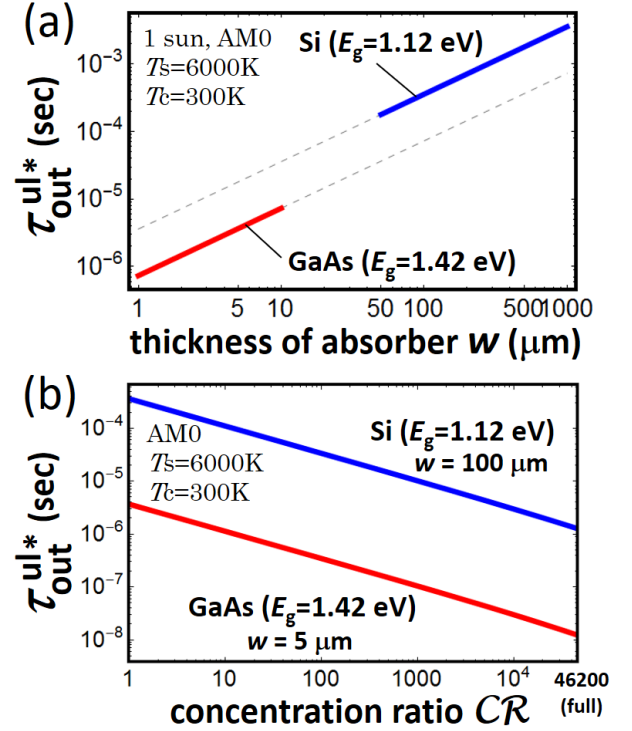}
\caption{Upper boundary of the extraction time $\tau_{\rm out}^{{\rm ul} \ast}$ when evaluated as 10 percent of $\tau_{\rm out}^{\rm ul}(E_e=k_B T_c)$ in Eq.~(\ref{eq:tau_out-UL1}), shown as functions of (a) cell thickness $w$ and (b) the concentration ratio for planar Si and GaAs solar cells. In (a), we assumed 1 sun illumination ($\mathcal{CR}=1$) and maximum power operation at $V=V_{\rm mp}$ (=0.779 V for Si and 1.052 V for GaAs). In (b), the absorber thickness $w$ was fixed at $100 \mu$m for Si and $5 \mu$m for the GaAs cells. The $\mathcal{CR}$-dependence of $V_{\rm mp}(\mathcal{CR})$ is calculated using the SQ theory.}  
\label{fig9} 
\end{figure}

In Fig.~\ref{fig9}, we plotted the boundary time $\tau_{\rm out}^{{\rm ul} \ast}$ that was defined using 10 percent of $\tau_{\rm out}^{\rm ul}(E_e=k_B T_c)$ as a function of absorber thickness $w$ for Si and GaAs solar cells.
Fig.~\ref{fig9}(a) clearly shows that $\tau_{\rm out}^{{\rm ul} \ast}$ is dependent on the material parameters (effective mass and indirect/direct gap) and is linearly on $w$. For 100-$\mu$m-thick Si solar cells, $\tau_{\rm out}^{{\rm ul} \ast}$ is estimated to be on the sub-ms scale. 
Additionally, $\tau_{\rm out}^{{\rm ul} \ast}$ will decrease in the concentrator solar cells and is plotted as a function of concentration ratio $\mathcal{CR}$ in Fig.~\ref{fig9}(b). 
$\tau_{\rm out}^{{\rm ul} \ast}$ is roughly proportional to $1/\sqrt{\mathcal{CR}}$.

The physical meaning of this condition still seems unclear since $\tau_{\rm out}^{\rm ul}(E_{e(h)})$ in Eq.~(\ref{eq:tau_out-UL1}) is dependent on energies of the microscopic carrier states. 
The physical meaning of the timescale will become clearer if the condition $\Delta n^{e(h)}_{E_{e(h)}} \ll  f^{F}_{\mu_{e(h)}, \beta_c}(E_{e(h)})$ is summed over all the microscopic states.
This can be read as 
\begin{eqnarray}
\tau_{\rm out}  &\ll &\frac{ \mathcal{V} \int \mathcal{D}_{e(h)}(E_{e(h)}) f^{F}_{\mu_{e(h)}, \beta_c}(E_{e(h)}) {\rm d}E_{e(h)}}{\mathcal{V} \int \mathcal{D}_{e(h)}(E_{e(h)}) J^{e(h),{\rm sun}}_{E_{e(h)}}{\rm d}E_{e(h) } }  \nonumber \\
&\approx & \frac{N_c}{I_{\rm sc}^{\rm max} /|e|} \left(=\frac{n_c}{i_{\rm sc}^{\rm max}/|e|} \times w \right), \label{eq:tau_out-UL2}
\end{eqnarray}
where $N_c(=N_e=N_h)$ is the total number of carriers, $n_c \equiv N_c/\mathcal{V}$ is the carrier density (where $n_c w = N_c/\mathcal{A}$ is the areal density); the maximum available short-circuit current $I_{\rm sc}^{\rm max}$ is given by
\begin{eqnarray}
I_{\rm sc}^{\rm max} \equiv |e| \times \mathcal{A} \times \int_{E_g}^{\infty}  j^{\rm sun}(E) {\rm d}E \label{eq:MaxIsc},
\end{eqnarray}
from which the corresponding current density per unit area is defined using 
\begin{eqnarray}
i_{\rm sc}^{\rm max} =I_{\rm sc}^{\rm max} /\mathcal{A}.
\end{eqnarray}
The meaning of the condition in Eq.~(\ref{eq:tau_out-UL2}), which is equivalent to $N_c/\tau_{\rm out} \gg I_{\rm sc}^{\rm max} /|e|$, is now much clearer.
The number of carriers output from the absorber per unit time (which differs from the net current given by the outflow minus the inflow) must be greater than the number of photons absorbed per unit time, i.e., the number of carriers that is generated per unit time in the absorber.
This condition may have simply been assumed in the SQ theory, although it is not given explicitly in their original paper~\cite{SQ}.  

The final equation in Eq.~(\ref{eq:tau_out-UL2}) clearly explains the $w$-linear dependence of $\tau_{\rm out}^{{\rm ul} \ast }$ shown in Fig.~\ref{fig9}(a).
Note here that $i_{\rm sc}^{\rm max}$ and the carrier density $n_c$ are independent of $w$ because $i_{\rm sc}^{\rm max}$ is given solely by the Sun illumination conditions and $n_c$ is given by the chemical potential that is fixed at the maximum operating power point, $V=V_{\rm mp}$. 
The $\mathcal{CR}$-dependence of $\tau_{\rm out}^{{\rm ul}\ast} \propto 1/\sqrt{\mathcal{CR}}$ shown by Fig.~\ref{fig9}(b) can also be understood based on the following analysis. The radiative loss current $I_{\rm rad}$ at the maximum operating power can be estimated from the SQ theory using the $I$-$V$ relation, where $I=I_{\rm sun}-I_{\rm rad}$ with $I_{\rm rad}=I_{\rm rad}^0 e^{\beta_c |{\rm e}|V}$ (see Sec.~\ref{SecI}). 
Using ${\rm d}(IV)/{\rm d}V=0$, we can easily show that $I_{\rm rad} \approx I_{\rm sun}/(\beta_c V_{\rm mp})$ as long as $\beta_c V_{\rm mp} \gg 1$, which is normally fulfilled. 
In general, $V_{\rm mp}$ increases with the concentration ratio $\mathcal{CR}$; at the same time, however, it can never exceed the absorber band gap $E_g$ for solar cell operation (below the lasing condition~\cite{Bernard}), i.e., $V_{\rm mp}(\mathcal{CR}=1)\le V_{\rm mp}(\mathcal{CR}) \le E_g$.
In Si, for example, this means that $0.779 \ {\rm eV} \le V_{\rm mp}(\mathcal{CR}) \le 1.12 \ {\rm eV}$, and correspondingly, $I_{\rm rad} \left( \approx I_{\rm sun}/(\beta_c V_{\rm mp}) \right)$ ranges at most from 2.3 to 3.3 percent of $I_{\rm sun}$ over the whole $\mathcal{CR}$ range $(1 \le  \mathcal{CR} \le 46200 \ ({\rm full}))$. 
Therefore, $I_{\rm rad}$ is almost proportional to $I_{\rm sun} (\propto \mathcal{CR})$.
In addition, we notice that the radiative loss current in general is $I_{\rm rad} \propto n_c^2$, or equivalently, $n_c \propto I_{\rm rad}^{1/2} \propto \mathcal{CR}^{1/2}$. Because $i_{\rm sc}^{\rm max}$ is obviously $\propto I_{\rm sun} \propto \mathcal{CR}$, another definition of $\tau_{\rm out}^{{\rm ul}\ast}$ from Eq.~(\ref{eq:tau_out-UL2}) shows that $\tau_{\rm out}^{{\rm ul}\ast} \propto n_c/i_{\rm sc}^{\rm max} \propto\mathcal{CR}^{-1/2} $.

\section{\label{SecIV} Numerical simulation of device performance and conversion efficiency limit of a simple planar solar cell}

\begin{figure}[!t]
\centering
\includegraphics[width=0.49\textwidth]{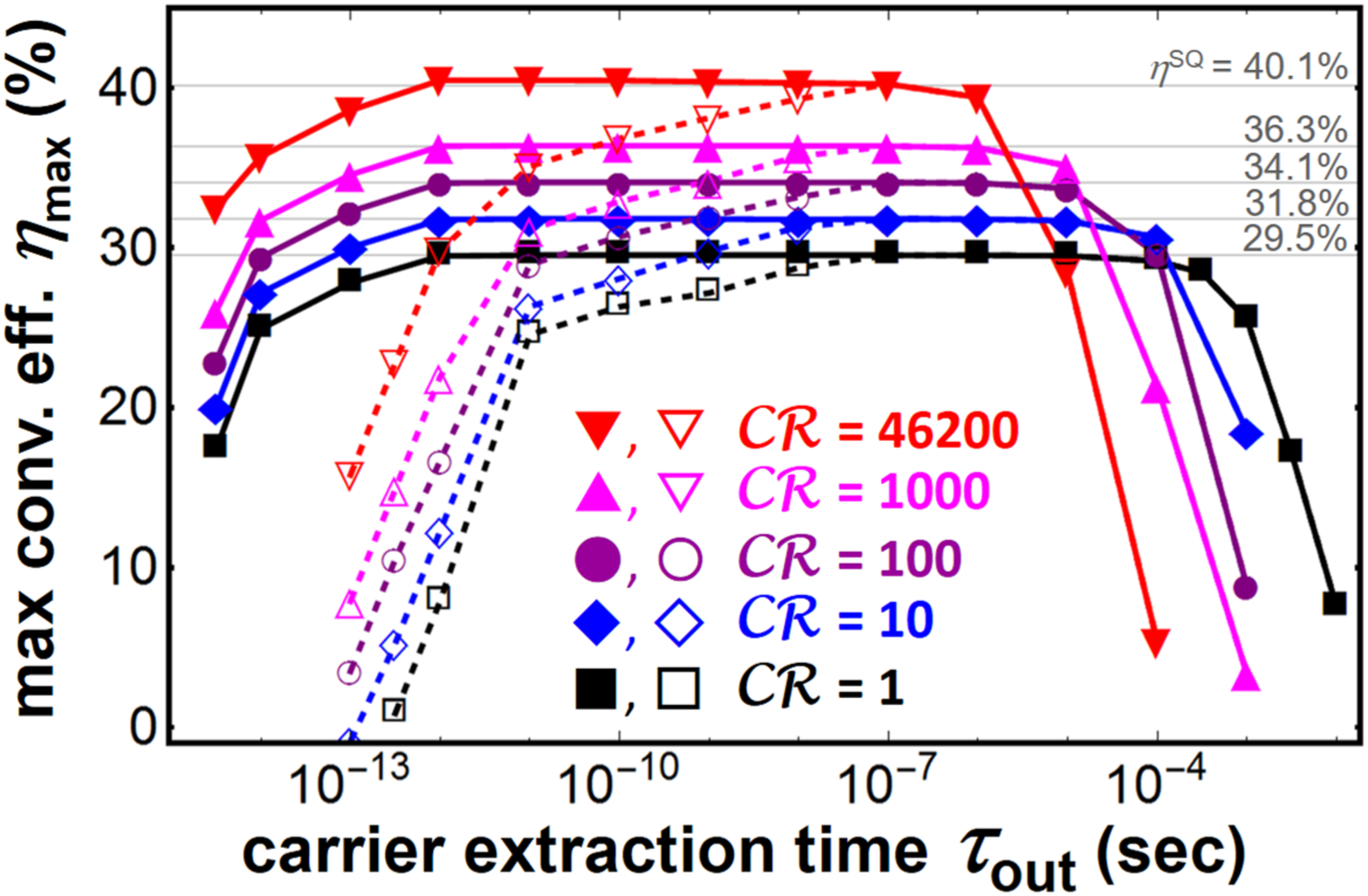}
\caption{Summary of device simulation results for simple planar solar cell with a $100-\mu$m-thick Si absorber using our nonequilibrium theory; the maximum conversion efficiency is shown as a function of carrier extraction time $\tau_{\rm out}$ for various concentration ratios ($\mathcal{CR}=1, \ 10^1, \ 10^2, \ 10^3,  \  46200$).
The solid lines were obtained for the heat-shared phonon reservoirs and the dashed lines were obtained for the heat-isolated phonon reservoirs (as explained in the text). The realistic limit for the maximum efficiency will lie between the solid and dashed curves. The SQ limit for each $\mathcal{CR}$ is also indicated using horizontal lines.}  
\label{fig10} 
\end{figure}

\begin{figure}[!t] 
\centering
\includegraphics[width=0.49\textwidth]{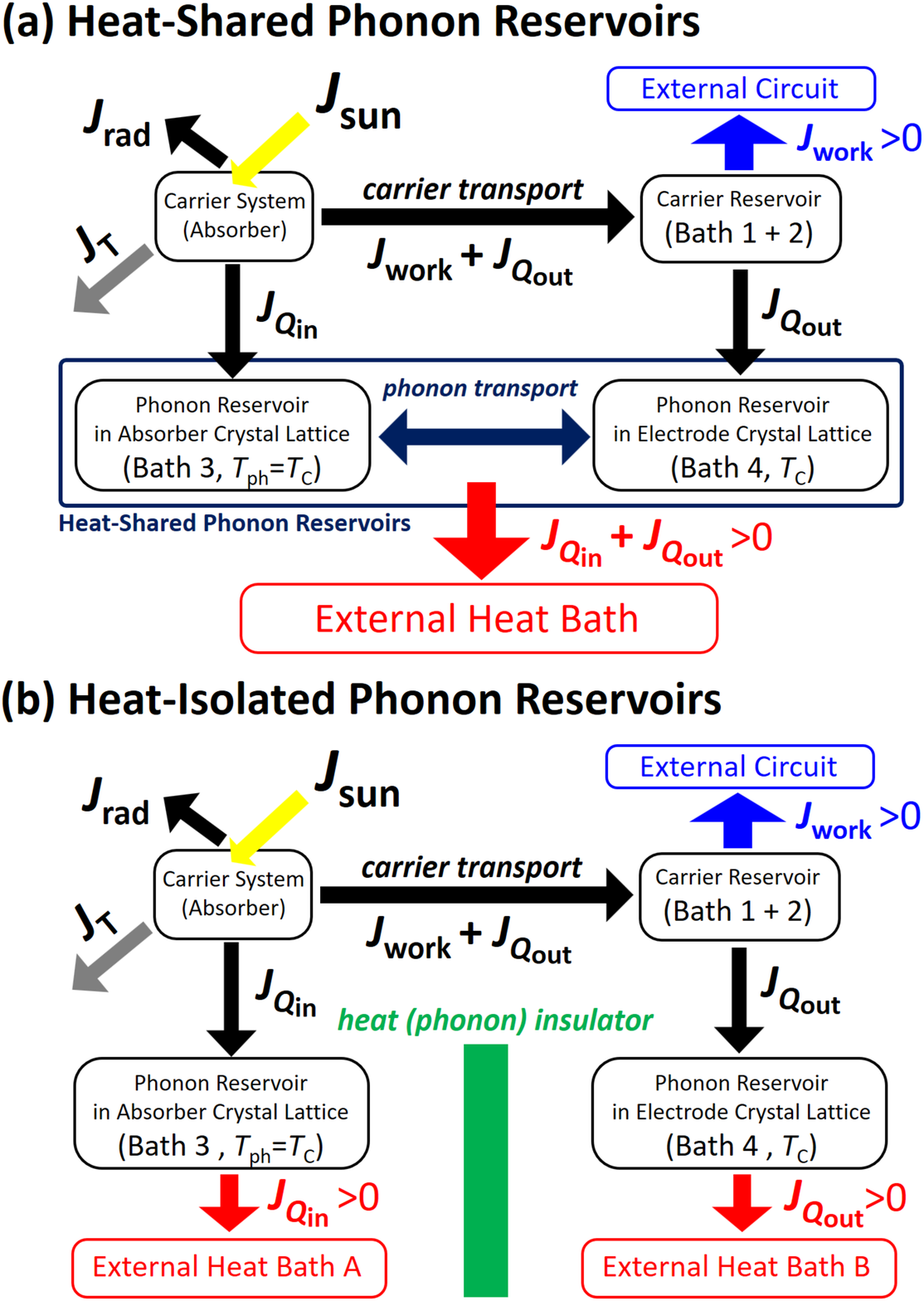}
\caption{Two ideal models that take the phonon environments in the two limiting cases into account: (a) heat-shared phonon reservoirs, where the phonon transport between the absorber and electrodes is very fast, in either direct or indirect ways; and (b) heat-isolated phonon reservoirs, where the phonon transport between the absorber and electrodes is very slow. In the latter case, the heat generated in the absorber ($J_{Q_{\rm in}}$) and in the electrodes ($J_{Q_{\rm out}}$) are dissipated independently into different heat baths (heat baths A and B).
TO enable solar cell operation, $J_{\rm work}>0$ must be self-sustained solely by solar illumination $J_{\rm sun}>0$; the requirements are that $J_{Q_{\rm in}}+J_{Q_{\rm out}}>0$ for (a) heat-shared phonon reservoirs, and that $J_{Q_{\rm in}}>0$ and $J_{Q_{\rm out}}>0$ for (b) the heat-isolated phonon reservoirs. Otherwise, an additional external heat supply to the absorber and/or the electrodes will be required to sustain solar cell operation; the definition of the solar cell conversion efficiency then becomes less clear. }  
\label{fig11}  
\end{figure}

In this section, we present numerical simulation results for the device performance and the conversion efficiency limit of a simple planar solar cell.
The results obtained here are summarized in Fig.~\ref{fig10}, which shows the theoretical conversion efficiency limit as a function of carrier extraction time $\tau_{\rm out}$ for various values of concentration ratio $\mathcal{CR}$. 
As shown in Fig.~\ref{fig10}, two curves (solid and dashed lines) are presented for a given $\mathcal{CR}$. 
The theoretical limit is dependent on how effectively the heat that is generated in the crystal lattices of the absorber and the electrodes can be delivered to each other via phonon transport.
In the ideal limit case, where phonon transport between the absorber and the electrodes is very fast in either direct or indirect ways, which we shall refer to as ``heat-shared phonon reservoirs '' (see Fig.~\ref{fig11}(a)), we have higher maximum conversion efficiencies (shown as solid curves in Fig.~\ref{fig10}) when the carrier extraction becomes fast.
At the opposite limit, where the phonon transport between absorber and electrodes so slow as to be negligible, which we refer to as ``heat-isolated phonon reservoirs'' (Fig.~\ref{fig11}(b)), the maximum conversion efficiency is reduced when the carrier extraction becomes fast (see the dashed curves in Fig.~\ref{fig10}).
In realistic cases, the maximum conversion efficiency will lie somewhere between these two curves. The precise position will be dependent on the phonon transport properties between the absorber and the electrodes, which will be sensitive to the formation conditions for the contacts between the absorber and the electrodes and/or the phonon environment surrounding them.

The differences in the two ideal cases are incorporated into the stability conditions for solar cell operation. 
To enable solar cell operation with $J_{\rm work}>0$ to be self-sustained solely by solar illumination with $J_{\rm sun}>0$, the requirements are that $J_{Q_{\rm in}}+J_{Q_{\rm out}}>0$ for the heat-shared phonon reservoirs shown in Fig.~\ref{fig11}(a), and that $J_{Q_{\rm in}}>0$ and $J_{Q_{\rm out}}>0$ for the heat-isolated phonon reservoirs shown in Fig.~\ref{fig11}(b). 
Otherwise, an additional external heat supply to the absorber and/or the electrodes will be required to sustain solar cell operation, and the definition of the solar cell conversion efficiency then becomes less clear.
In the following subsections, we discuss the solar cell properties and the underlying physics for these two limiting cases.

\subsection{\label{SecIV-A} A limiting case: heat-shared phonon reservoirs}
We now present numerical simulation results for a limiting case involving heat-shared phonon reservoirs.
The solid lines in Fig.~\ref{fig10} show the maximum conversion efficiencies for various concentration ratios ($\mathcal{CR}=1,10,10^2,10^3, 46200$) plotted as a function of the carrier extraction time $\tau_{\rm out}$.
We find flat regions for $\tau_{\rm out}$ of more than 1 ps and less than $\tau_{\rm out}^{\rm ul \ast}$, which is dependent on the concentration ratio $\mathcal{CR}$ (Fig.~\ref{fig9} (b)), while the maximum conversion efficiency is equal to the SQ limit $\eta^{\rm SQ}$.
As expected, we can conclude that the SQ theory applies in the normal extraction regime II.
Outside this regime, in both the slow and fast extraction regimes denoted by I and III, respectively, we found significant reductions in $\eta_{\rm max}$ for the simple planar solar cell (Fig.~\ref{fig3}).
Because the solar cell properties are different for each regimes, our definition of the parameter classifications in Fig.~\ref{fig8} seems reasonable. 
We must now consider how the difference can be understood.

\begin{figure}[!t]
\centering
\includegraphics[width=0.47\textwidth]{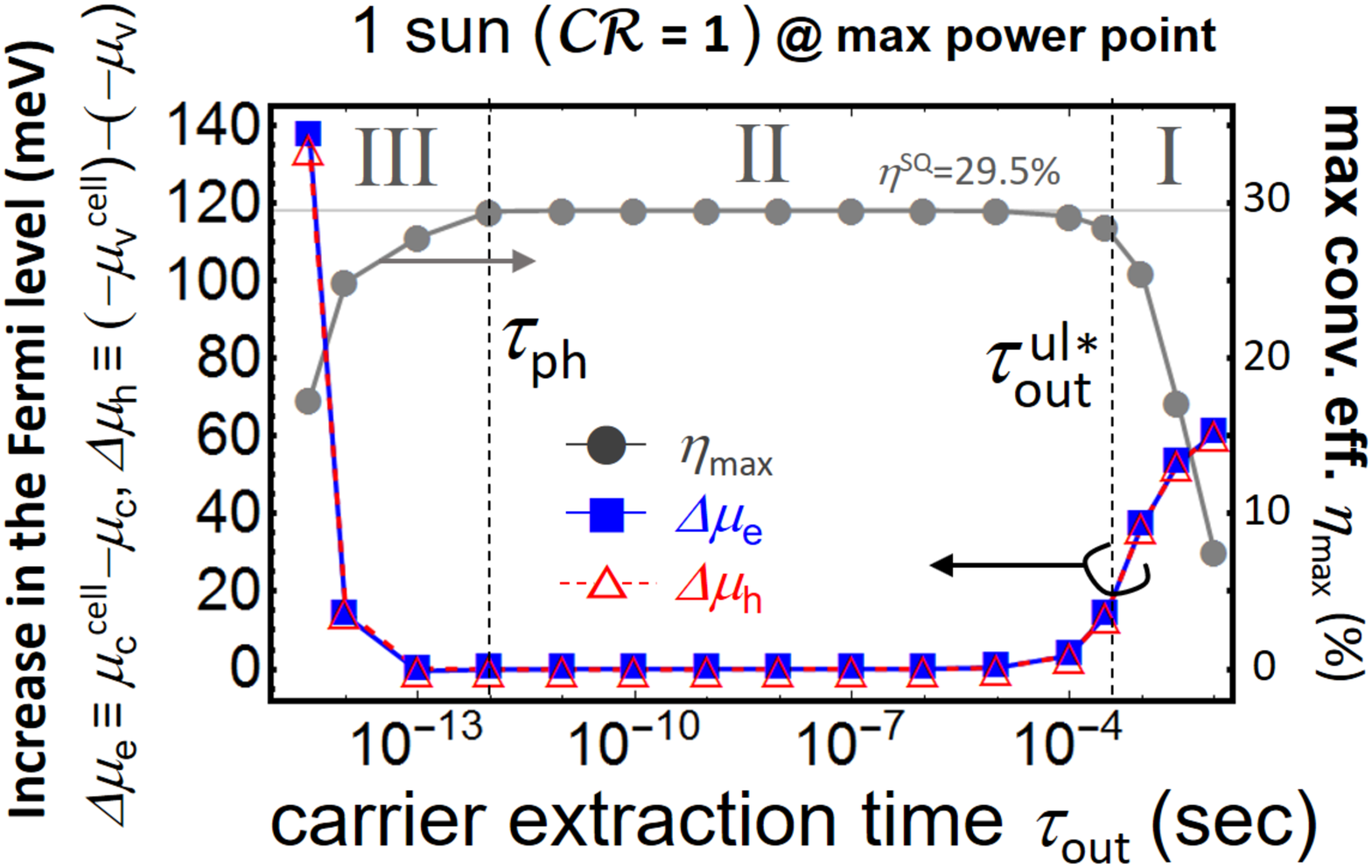}
\caption{Increases in the Fermi levels of electrons and holes in the absorber when compared with that in the electrodes; $\Delta \mu_e \equiv \mu_c^{\rm cell}-\mu_c$ and $\Delta \mu_h \equiv (-\mu_v^{\rm cell})-(-\mu_v)$ are evaluated at the maximum operating power point $V=V_{\rm mp}$ and are shown as functions of carrier extraction time (for 100-$\mu$m-thick Si under 1 sun AM0 illumination). The maximum conversion efficiency (the black solid line shown in Fig.~\ref{fig10}) is also shown.}  
\label{fig12} 
\end{figure}

Fig. \ref{fig12} shows an increase in the Fermi levels of the electrons and holes in the absorber when compared with those in the electrodes, denoted by $\Delta \mu_e \equiv \mu_c^{\rm cell}-\mu_c$ and $\Delta \mu_h \equiv (-\mu_v^{\rm cell})-(-\mu_v)$, respectively, when evaluated using the carrier distribution functions at the maximum operating power point $V=V_{\rm mp}$ under 1 sun illumination ($\mathcal{CR}=1$). 
The increases in the Fermi levels found in regimes I and III clearly show that the carriers in the absorber form nonequilibrium populations.
We checked numerically that similar results were also obtained for higher values of concentration ratio $\mathcal{CR}>1$.
The increased Fermi levels in the different regimes originate from different mechanisms.
In regime I, the carriers are accumulated in the absorber because of the slow extraction process, resulting in increases in the carrier density and Fermi level.
In contrast, the increment in regime III is attributed to broadening of the microscopic states because $\Gamma=\hbar / \left(2 \sqrt{\log{2}} \tau_{\rm out} \right)$ is no longer negligible.
In this regime, the dominant term in the rate equation is the carrier extraction term $J^{e(h),{\rm out}}_{E_{e(h)}}$, which allows us to have $J^{e(h),{\rm out}}_{E_{e(h)}}\approx 0$. 
As a result, the carrier distribution functions in the absorber can be approximated using $ n^e_{E_e}\approx \tilde{f}^{F}_{\mu_c, \beta_c}(E_g+E_e)$ and $n^h_{E_h}\approx \tilde{f}^{F}_{-\mu_v, \beta_c}(E_h)$ from Eq.~(\ref{eq:Jeout-renew}) and Eq.~(\ref{eq:Jhout-renew}). 
As long as the band filling effect remains negligible, i.e., when $n^e_{E_e}\ll1$ and $n^h_{E_h} \ll 1$, the distribution functions from Eq.~(\ref{eq:ConvolvedFDfunc}) can be approximated using
\begin{eqnarray}
n^e_{E_e}&\approx& \exp{ \left( -\beta_c (E_g+E_e-(\mu_c +\beta_c \Gamma^2/4) \right)}, \\
n^h_{E_h}&\approx& \exp{\left(-\beta_c (E_h-(-\mu_v +\beta_c \Gamma^2/4) \right)}.
\end{eqnarray} 
Therefore, in this regime, the increases in the Fermi levels are fitted well using
\begin{eqnarray}
\Delta \mu_{e} & \equiv& \mu_c^{\rm cell}-\mu_c \approx \beta_c \Gamma^2/4,  \label{eq: electron excess heat} \\
\Delta \mu_h & \equiv& (-\mu_v^{\rm cell})-(-\mu_v) \approx \beta_c \Gamma^2/4.  \label{eq: hole excess heat}
\end{eqnarray}

\begin{figure}[!t]
\centering
\includegraphics[width=0.47\textwidth]{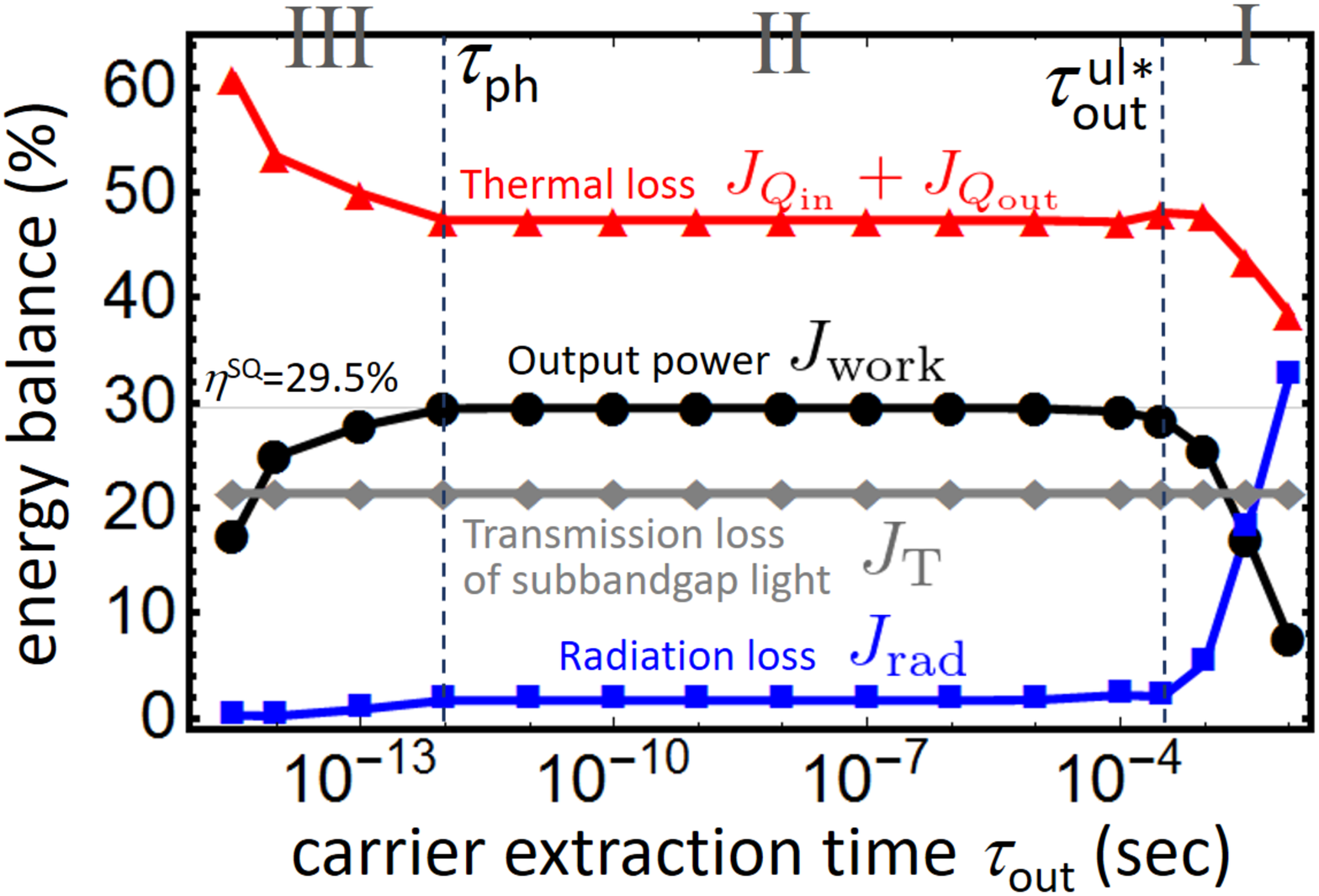}
\caption{Energy balance evaluated at the maximum operating power point $V=V_{\rm mp}$ as a function of carrier extraction time (100-$\mu$m-thick Si under 1 sun AM0 illumination). }  
\label{fig13}  
\end{figure}

To see what happened in the nonequilibrium regimes directly, we simulated the energy balance at the maximum operating power point, as shown in Fig.~\ref{fig13}, which provides further information on the energy loss mechanism.
In slow extraction regime I, the energy loss caused by radiative recombination ($J_{\rm rad}$) increases greatly with increasing $\tau_{\rm out}$, while that due to thermal loss ($J_{Q_{\rm in}}+J_{Q_{\rm out}}$) decreases. 
This can be understood easily because the slow extraction process increases the carrier density in the absorber, which thus enhances the radiative recombination rate. 
In contrast, in fast extraction regime III, the thermal loss increases while the radiation loss decreases as $\tau_{\rm out}$ decreases because the excess energy of the photo-generated carriers is transferred quickly and dissipated rapidly in the electrodes before thermalization in the absorber is complete.
In this sense, $\Delta \mu_{e}$ in Eq.~(\ref{eq: electron excess heat}) and $\Delta \mu_{h}$ in Eq.~(\ref{eq: hole excess heat}) can be regarded as additional excess heat energy conveyed by the fast extraction of one carrier. 
As shown in Fig.~\ref{fig13}, the requirement for stable solar cell operation, given by $J_{Q_{\rm in}}+J_{Q_{\rm out}}>0$ in this model with heat-shared phonon reservoirs, is fulfilled (and is also fulfilled for $0<V<V_{\rm oc}$, which is not shown here).

\begin{figure}[!t]
\centering
\includegraphics[width=0.47\textwidth]{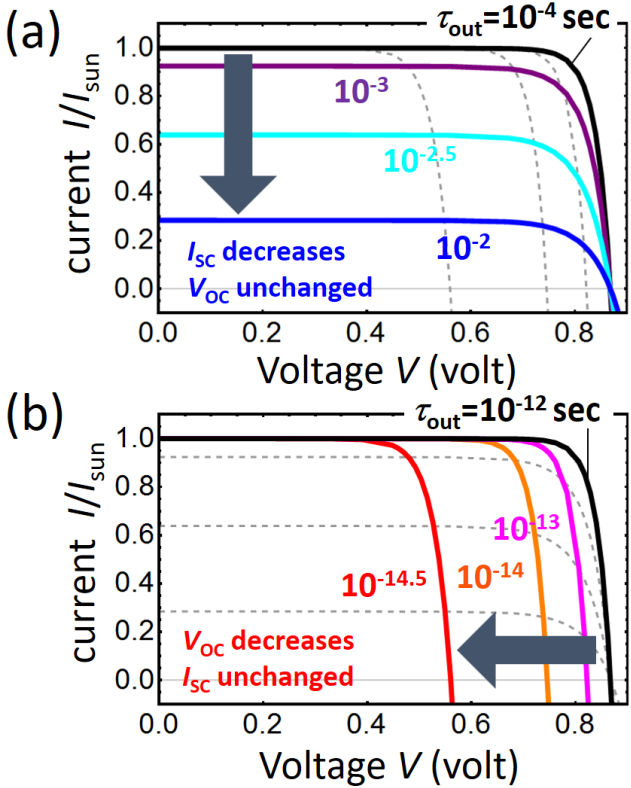}
\caption{Current-voltage characteristics for various carrier extraction times (100-$\mu$m-thick Si under 1 sun AM0 illumination): (a) curves obtained in slow extraction regime I are highlighted; (b) curves obtained in fast extraction regime III are highlighted. The solid black curves in (a) and (b) were obtained for $10^{-12}{\rm sec}<\tau_{\rm out}<10^{-4}{\rm sec}$ in normal extraction regime II.}  
\label{fig14}  
\end{figure}

The differences between these loss mechanisms are reflected in their current-voltage ($I$-$V$) characteristics in each regime, as shown in Fig.~\ref{fig14}.
In slow extraction regime I (Fig.~\ref{fig14}(a)), the short-circuit current $I_{\rm sc}$ decreases with increasing $\tau_{\rm out}$ while the open-circuit voltage $V_{\rm oc}$ remains unchanged. In contrast, in fast extraction regime III (Fig.~\ref{fig14}(b)), the open-circuit voltage $V_{\rm oc}$ decreases with decreasing $\tau_{\rm out}$ while the short-circuit current  $I_{\rm sc}$ remains unchanged. These results agree with the consideration that the enhanced radiation loss in regime I accompanies a loss in the output charge current while the enhanced heat current in regime III does not accompany such a loss.
We find no significant changes in the $I$-$V$ characteristics for $\tau_{\rm out}$ between 1 ps and $10^{-3.5}$ s, thus supporting the belief that the SQ theory works in normal extraction regime II.

The reduction of $\eta_{\rm max}$ in fast extraction regime III in Fig.~\ref{fig10} may be confusing because the hot carrier solar cells that are targeted in this regime can surpass the SQ limit of $\eta^{\rm SQ}$ theoretically~\cite{Nozik, Wurfel2, Takeda1, Takeda2, Suchet}. Therefore, we may expect an increase in $\eta_{\rm max}$ as $\tau_{\rm out}$ decreases in regime III.
However, this discrepancy is not surprising and can be explained as follows. 
The differences in the results stem from the differences between the carrier extraction processes.
A hot carrier solar cell uses a filter to select the energies of the carriers that pass from the absorber to the electrodes, which can reduce the heat dissipation (thermal losses) in the electrodes. Tailored filtering can increase the output voltage while preventing large output current losses in hot carrier solar cells.
However, our simple planar solar cell in Fig.~\ref{fig10} does not use such a filter, and the heat dissipation in the electrodes is therefore not controlled.
As already shown in Fig.~\ref{fig13} and as will be shown in the next subsection (Fig.~\ref{fig15}), heat losses, and especially the heat loss in the electrodes, 
increase in regime III. This results in a strong reduction in $\eta_{\rm max}$ in our case.
These contrasting results support the supposition that, unless a tailored carrier extraction process such as that using energy selection is used, it is difficult for fast carrier extraction before thermalization to be beneficial.

\subsection{\label{SecIV-B} Another limiting case: heat-isolated phonon reservoirs}
The maximum conversion efficiency $\eta_{\rm max}$ is also dependent on the phonon environment that surrounds the solar cell. 
In this subsection, we focus on solar cell performance in another limiting case with the heat-isolated phonon reservoirs shown in Fig.~\ref{fig11}(b). 
When exchange of phonons between the absorber and electrode crystals is prevented both directly and indirectly, i.e. when their phonon environments are isolated (e.g., Baths 3 and 4 in Fig.~\ref{fig11}(b)), $\eta_{\rm max}$ is significantly reduced from the values obtained for heat-shared reservoirs for a small $\tau_{\rm out}$.
Similar results are obtained, irrespective of the value of $\mathcal{CR}$.

Here we explain how these differences occur.
In the heat-isolated phonon reservoirs, the stability condition for the solar cell requires $J_{Q_{\rm in}}>0$ and $J_{Q_{\rm out}}>0$.
Otherwise, an additional heat supply must be added to the absorber or the electrodes, which is not suitable for our targeted device.
To be more precise, let's consider a situation where $J_{Q_{\rm out}}<0$ (as found below) and what will happen next in the device.
In this case, the electrodes require heat supply from others for the stable operation. However, supply from the absorber lattice is prohibited in the heat isolated model. With no heat supply, the lattice temperature of the electrodes, initially at the ambient temperature, will decrease. Then, the cooled electrodes will start to collect heat from the ambient (e.g. surrounding air) at the higher temperature. 
Then, the cooled electrodes will cool the ambient next during the solar cell operation. 
Semiconductor devices with similar structure, which cool down the ambient for the operation, can exist in reality as found in light-emitting diodes (LEDs) (namely, refrigerating LEDs~\cite{Xue, Tauc, Dousmanis, Berdahl, Santhanam}). 
However solar cells are the devices aimed for long-term stand-alone operation requiring the stability at a long-time scale.
In our model simulation with heat isolated reservoirs, we consider the device with $J_{Q_{\rm out}}<0$ inevitably requires an additional heat supply for the stable operation.

\begin{figure}[!t] 
\centering
\includegraphics[width=0.49\textwidth]{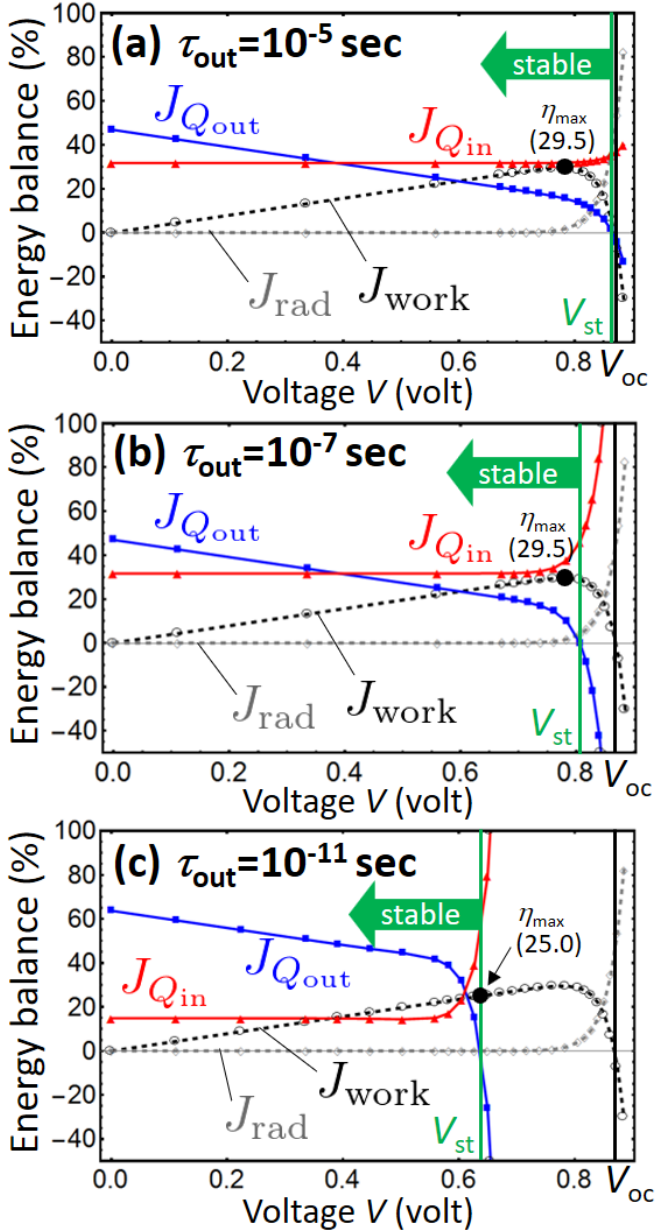}
\caption{Energy balance shown as a function of bias voltage $V$ for various carrier extraction times: (a) $\tau_{\rm out}=10^{-5}$, (b) $10^{-7}$, and (c) $10^{-11}$ s (100-$\mu$m-thick Si under 1 sun AM0 illumination). For the ideal model with heat-isolated phonon reservoirs shown in Fig.~\ref{fig11}(b), the stability condition is given by $J_{Q_{\rm in}}>0$ and $J_{Q_{\rm out}}>0$, which is fulfilled for $V<V_{\rm st}$ (green vertical lines). Note that the fraction from transmission loss $J_{\rm T}$ of the subbandgap light is 21.4 percent (not shown here). }  
\label{fig15}  
\end{figure}

In Fig.~\ref{fig15}, the energy balance is shown as a function of bias voltage $V$ for various carrier extraction times: (a) $\tau_{\rm out}=10^{-5}$, (b) $10^{-7}$, and (c) $10^{-11}$ s.
To address the stability condition, we plotted the contributions from the heat flow into the absorber $J_{Q_{\rm in}}$ and into the electrodes $J_{Q_{\rm out}}$ separately.
As shown by the plots, $J_{Q_{\rm out}}$ decreases with $V$ and changes sign from positive to negative at a point $V=V_{\rm st}$, whereas $J_{Q_{\rm in}}$ always remains positive. 
We call $V_{\rm st}$ the stability boundary here because the stability condition is fulfilled for $V<V_{\rm st}$.

When the carrier extraction is slow, e.g. when $\tau_{\rm out}=10^{-5}$ s in Fig.~\ref{fig15}(a), $V_{\rm st}$ is almost the same as $V_{\rm oc}$.
This means that the heat flow direction between absorber and electrodes is the same as the charge flow direction.
We consider this to be the normal situation.
However, when the extraction becomes faster, as shown in Fig.~\ref{fig15}(b) and Fig.~\ref{fig15}(c), we find a clear departure of $V_{\rm st}$ from $V_{\rm oc}$. In this case, we found the regime where $V_{\rm st}<V<V_{\rm oc}$, in which $J_{\rm work}>0$ but $J_{Q_{\rm out}}<0$, i.e., where the heat flow and charge flow directions are different.
In this regime, the solar cell generates electric power ($J_{\rm work}>0$), but an external heat supply of no less than $|J_{Q_{\rm out}}|$ must be additionally provided for the electrodes.
Such a device would not provide a solar cell operating in a self-sustained manner using solar illumination alone.
In this way, we have evaluated $\eta_{\rm max}$ to be the maximum conversion efficiency under the stability conditions $0<V<V_{\rm st}$. 
For example, in Fig.~\ref{fig15}, $\eta_{\rm max}$ is 29.5 percent for $\tau_{\rm out}=10^{-5}$ and $10^{-7}$ s, whereas $\eta_{\rm max}$ is reduced to 25.0 percent for $\tau_{\rm out}=10^{-11}$ s.

As already shown in Fig.~\ref{fig13}, the sum $J_{Q_{\rm in}}+J_{Q_{\rm out}}$ is positive as long as $0<V<V_{\rm oc}$.
Therefore, if thermally linked, the depleted heat in the electrodes when $V_{\rm st}<V<V_{\rm oc}$ can be complemented by the heat inflow into the absorber denoted by $J_{Q_{\rm in}}$. 
The ideal limit in such a case with a strong thermal link corresponds to the heat-shared phonon reservoirs that were discussed in Sec.~\ref{SecIV-A}.
In real situations where the thermal link strength is moderate (i.e., heat depletion in the electrodes is partially complemented by the absorber via phonon transport), $\eta_{\rm max}$ will be located between the two ideal cases (the solid and dashed curves in Fig.~\ref{fig10}).
Because the difference between these two results is large, we propose that $\eta_{\rm max}$ is sensitive to the phonon transport between the absorber and the electrodes (in either direct or indirect ways) in the fast carrier extraction regime.

\section{\label{SecV} Conclusion and future prospects}
We have formulated a nonequilibrium theory for solar cells that includes microscopic descriptions of the carrier thermalization and extraction processes.
This theory extends the Shockley-Queisser theory to nonequilibrium parameter regimes where the detailed balance cannot be applied.
The theory provides detailed information about the solar cells, including nonequilibrium carrier distribution functions with the chemical potentials that are higher than those in the electrodes, and the energy balance (including output work, radiation losses, transmission losses, and heat dissipation in the absorber and the electrodes), which will provide a precise understanding of the loss mechanisms in various solar cell types for a wide range of parameters. 

Using the developed theory, we defined three different regimes in terms of their carrier extraction time that were bounded using two time scales: the thermalization time, $\tau_{\rm ph}$, and $\tau_{\rm out}^{{\rm ul} \ast}$, at which the device characteristics should change. 
The upper boundary $\tau_{\rm out}^{{\rm ul} \ast}$ is dependent on the absorber material parameters, and is more strongly dependent on the system parameters, e.g., the absorber thickness and solar light concentration ratio. 

Device simulations of simple planar solar cells have shown that the SQ limit is applicable in the normal extraction regime, denoted by regime II ($\tau_{\rm ph}<\tau_{\rm out}<\tau_{\rm out}^{{\rm ul} \ast}$) in Fig.~\ref{fig8}.
Outside this regime (in regimes I and III), nonequilibrium carrier populations are found in the absorber and the maximum conversion efficiency is significantly reduced from the SQ limit.
While the reductions in $\eta_{\rm max}$ were similar, the energy loss mechanisms in the fast and slow extraction regimes are different, which is clearly reflected in their $I$-$V$ characteristics.
The reduction in $\eta_{\rm max}$ in the fast extraction regime also indicates that unless a tailored carrier extraction procedure such as that based on energy selection was performed, it would be difficult for fast carrier extraction before carrier thermalization to be beneficial.
This strong claim is consistent with the fact that hot carrier solar cells require energy selection during their carrier extraction processes in addition to the fast extraction procedure.

The nonequilibrium theory presented here covers only a few basic elements of solar cells and has only been tested in simple planar solar cells.
The losses of photo-generated carriers in the absorber in this work are solely due to radiative recombination. 
Inclusion of nonradiative recombination may change the result, as will be discussed elsewhere.
In the carrier extraction process, this paper does not consider energy losses at the junction. 
A case of this type using ohmic contacts will be studied in future work.
Application of the proposed theory to other types of solar cells, e.g., organic solar cells, perovskite solar cells, multi-junction solar cells, intermediate-band solar cells, and hot carrier solar cells will also be interesting.
The most important and challenging aspect will be to provide feasible proposals for new solar cells using nonequilibrium features by which the SQ limit can be surpassed.

\acknowledgements
Authors acknowledge Tatsuro Yuge, Makoto Yamaguchi, Yasuhiro Yamada, Katsuhiko Shirasawa, Tetsuo Fukuda, Katsuhito Tanahashi, Tomihisa Tachibana, and Yasuhiko Takeda for discussion.
This work is by JSPS KAKENHI (15K20931), and the New Energy and Industrial Technology Development Organization (NEDO).


\begin{thebibliography}{99}
\bibitem{SQ}
W.~Shockley, and H.~J.~Queisser, Detailed balance limit of efficiency of p-n junction solar cells, J. Appl. Phys. {\bf 32},  510 (1961).
\bibitem{Richter} 
A.~Richter, M.~Hermle, and S.~W.~Glunz, Reassessment of the Limiting Efficiency for Crystalline Silicon Solar Cells, IEEE J. Photovoltaics {\bf 3},  1184 (2013).
\bibitem{Nozik} R.~T.~Ross, and A.~J.~Nozik, Efficiency of hot-carrier solar energy converters, J. Appl. Phys. {\bf 53},  3813 (1982).
\bibitem{Wurfel2} P. W\"urfel, Solar energy conversion with hot electrons from impact ionisation, Sol. Energy Mater. Sol. Cells {\bf 46},  43 (1997).
\bibitem{Takeda1} Y. Takeda, T. Motohiro, D. K\"onig, P. Aliberti, Y. Feng, S. Shrestha, and G. Conibeer, Practical Factors Lowering Conversion Efficiency of Hot Carrier Solar Cells, Appl. Phys. Exp. {\bf 3},  104301 (2010).
\bibitem{Takeda2} Y. Takeda, A. Ichiki, Y. Kusano, N. Sugimoto, and T. Motohiro, Resonant tunneling diodes as energy-selective contacts used in hot-carrier solar cells, J. Appl. Phys. {\bf 118},  124510 (2015).
\bibitem{Suchet} D. Suchet, Z. Jehl, Y. Okada, and J-F. Guillemoles, Influence of Hot-Carrier Extraction from a Photovoltaic Absorber: An Evaporative Approach, Phys. Rev. Applied {\bf 8}, 034030 (2017).
\bibitem{EffTab50} M.~A.~Green, Y.~Hishikawa, W.~Warta, E.~D.~Dunlop, D.~H.~Levi, J.~Hohl-Ebinger, A.~W.~Ho-Baillie, Solar Cell Efficiency Tables (Version 50)'' Prog. Photovoltaics {\bf 25},  668 (2017).
\bibitem{Martin-Schwinger} P. C. Martin, and J. Schwinger, Theory of Many-Particle Systems. I, Phys. Rev. {\bf 115},  1342 (1959).
\bibitem{Kadanof-Baym} G. Baym, and Leo P. Kadanoff, Conservation Laws and Correlation Functions, Phys. Rev. {\bf 124},  287 (1961).
\bibitem{Keldysh} L. V. Keldysh, Diagram Technique for Nonequilibrium Processes, Sov. Phys.---JETP {\bf 20},  1018 (1965).
\bibitem{Rammer} J. Rammer, and H. Smith, Quantum field-theoretical methods in transport theory of metals, Rev. Mod. Phys. {\bf 58},  323 (1986).
\bibitem{Datta} S. Datta, {\it Electronic Transport in Mesoscopic Systems} (Cambridge University Press, Cambridge, 1995).
\bibitem{Steiger} S. Steiger, R. G. Veprek, and B. Witzigmann, Electroluminescence from a Quantum-Well LED using NEGF, in {\it Proceedings of the 13th International Workshop on ComputationalElectronics}, IWCE 2009, Beijing, China (IEEE 2009).
\bibitem{Henneberger} K. Henneberger, and H. Haug, Nonlinear optics and transport in laser-excited semiconductors,  Phys. Rev. B {\bf 38},  9759 (1988).
\bibitem{Lee} S. -C. Lee, and A. Wacker, Nonequilibrium Green's function theory for transport and gain properties of quantum cascade structures, Phys. Rev. B {\bf 66},  245314 (2002).
\bibitem{Szymanska} M. H. Szyma\'{n}ska, J. Keeling, and P. B. Littlewood, Nonequilibrium Quantum Condensation in an Incoherently Pumped Dissipative System, Phys. Rev. Lett. {\bf 96},  230602 (2006).
\bibitem{Yamaguchi} M. Yamaguchi, K. Kamide, R. Nii, T. Ogawa, and Y. Yamamoto, Second Thresholds in BEC-BCS-Laser Crossover of Exciton-Polariton Systems, Phys. Rev. Lett.  {\bf 111},  026404 (2013).
\bibitem{Aeberhard1} U. Aeberhard,   Quantum-kinetic theory of photocurrent generation via direct and phonon-mediated optical transitions, Phys. Rev. B {\bf 84},  035454 (2011).
\bibitem{Aeberhard2} U. Aeberhard,  Theory and simulation of quantum photovoltaic devices based on the nonequilibrium Green's function formalism, J. Comp. Electronics {\bf 10}, 394 (2011).
\bibitem{Cavassilas} N. Cavassilas, F. Michelini, and M. Bescond,  Modeling of nanoscale solar cells: The Green's function formalism, J. Renewable and Sustainable Energy {\bf 6},  011203 (2014).
\bibitem{Aeberhard3}  U. Aeberhard, and U. Rau,  Microscopic Perspective on Photovoltaic Reciprocity in Ultrathin Solar Cells, Phys. Rev. Lett. {\bf 118},  247702 (2017).
\bibitem{Rau} U. Rau,  Reciprocity relation between photovoltaic quantum efficiency and electroluminescent emission of solar cells, Phys. Rev. B {\bf 76},  085303 (2007).
\bibitem{Kirchartz} T. Kirchartz, and U. Rau,  Electroluminescence analysis of high efficiency Cu(In,Ga)Se2 solar cells, J. Appl. Phys.  {\bf 102},  104510 (2007).
\bibitem{Bernardi} M. Bernardi, D. Vigil-Fowler, J. Lischner, J. B. Neaton, and S. G. Louie,  {\it Ab Initio} Study of Hot Carriers in the First Picosecond after Sunlight Absorption in Silicon, Phys. Rev. Lett. {\bf 112},  257402 (2014).
\bibitem{Cardona} P. Y. Yu, and M. Cardona,  {\it Fundamentals of Semiconductors: Physics and Material Properties}, 3rd ed. (Springer, New York, 2005).
\bibitem{Wurfel1} P. W\"{u}rfel,  The chemical potential of radiation, J. Phys. C: Solid State Phys. {\bf 15},  3967 (1982).
\bibitem{Wurfel text} P. W\"{u}rfel, and U. W\"{u}rfel,  {\it Physics of Solar Cells: From Basic Principles to Advanced Concepts}, 3rd ed., (Weinheim: Wiley-VCH, 2016).
\bibitem{Carmichael} H. J. Carmichael,  {\it Statistical Methods in Quantum Optics 1: Master Equations and Fokker-Planck Equations}, 2nd ed.'' (Springer, 2003). 
\bibitem{Breuer} H. P. Breuer, and F. Petruccione,  The Theory of Open Quantum Systems, (Oxford University Press, 2002).
\bibitem{Tang} C. W. Tang,  Two-layer organic photovoltaic cell, Appl. Phys. Lett. {\bf 48},  183 (1986).
\bibitem{Heeger} G. Yu, J. Gao, J. C. Hummelen, F. Wudl, and A. J. Heeger,  Polymer Photovoltaic Cells: Enhanced Efficiencies via a Network of Internal Donor-Acceptor Heterojunctions, Science {\bf 270},  1789 (1995). 
\bibitem{Gratzel}  B. O'Regan and M. Gratzel,  A low-cost, high-efficiency solar cell based on dye-sensitized colloidal TiO$_2$ films, Nature {\bf 353},  737 (1991).
\bibitem{Miyasaka} A. Kojima, K. Teshima, Y. Shirai, and T. Miyasaka,  Organometal Halide Perovskites as Visible-Light Sensitizers for Photovoltaic Cells, J. Am. Chem. Soc. {\bf 131},  6050 (2009).
\bibitem{Frenkel} J. Frenkel,   On the Transformation of light into Heat in Solids. I, Phys. Rev. {\bf 37}, 17 (1931).
\bibitem{MarkFox} M. Fox, in Chapter 4 of  {\it Optical Properties of Solids}, 2nd ed., (Oxford University Press, 2010).
\bibitem{Ishii} Y. Tanaka, Y. Noguchi, K. Oda, Y. Nakayama, J. Takahashi, H. Tokairin, and H. Ishii,  Evaluation of internal potential distribution and carrier extraction properties of organic solar cells through Kelvin probe and time-of-flight measurements, J. Appl. Phys. {\bf 116},  114503 (2014).
\bibitem{Koster} V. M. Le Corre, A. R. Chatri, N. Y. Doumon, and L. J. A. Koster,  Charge Carrier Extraction in Organic Solar Cells Governed by Steady-State Mobilities, Adv. Energy Mater. {\bf 7},  1701138 (2017).
\bibitem{Suzuki} T. Suzuki, and R. Shimano,  Cooling dynamics of photoexcited carriers in Si studied using optical pump and terahertz probe spectroscopy, Phys. Rev. B {\bf 83},  085207 (2011).
\bibitem{Goldman} J. R. Goldman and J. A. Prybyla,  Ultrafast Dynamics of Laser-Excited Electron Distributions in Silicon, Phys. Rev. Lett.  {\bf 72},  1364 (1994).
\bibitem{Sabbah} A. J. Sabbah, and D. M. Riffe,  Femtosecond pump-probe reflectivity study of silicon carrier dynamics, Phys. Rev. B {\bf 66},  165217 (2002).
\bibitem{BookGreen} M. Green,  {\it Third Generation Photovoltaics: Advanced Solar Energy Conversion}, (Springer-Verlag: Berlin, Heidelberg, 2003).
\bibitem{Yablonovitch} E. Yablonovitch,  Statistical ray optics, J. Opt. Soc. Am.  {\bf 72},  899 (1982).
\bibitem{Chen} S. Chen, L. Zhu, M. Yoshita, T. Mochizuki, C. Kim, H. Akiyama, M. Imaizumi, and Y. Kanemitsu,  Thorough subcells diagnosis in a multi-junction solar cell via absolute electroluminescence-efficiency measurements, Sci. Rep.  {\bf 5},  7836 (2014).
\bibitem{Yu} J.-K. Yu, S. Mitrovic, D. Tham, J. Varghese, and J. R. Heath,  Reduction of thermal conductivity in phononic nanomesh structures, Nat. Nanotechnol.  {\bf 5},  718 (2010).
\bibitem{Hopkins} P. E. Hopkins, C. M. Reinke, M. F. Su, R. H. Olsson, E. A. Shaner, Z. C. Leseman, J. R. Serrano, L. M. Phinney, and I. El-Kady,   Reduction in the Thermal Conductivity of Single Crystalline Silicon by Phononic Crystal Patterning, Nano Lett.  {\bf 11},  107 (2011).
\bibitem{Bernard} M. G. A. Bernard, and G. Duraffourg,  Laser Conditions in Semiconductors, phys. stat. solidi b {\bf 1},  699 (1961).
\bibitem{Xue} J. Xue, Z. Li, and R. J. Ram, Irreversible Thermodynamic Bound for the Efficiency of Light-Emitting Diodes, Phys. Rev. Applied {\bf 8}, 014017 (2017). 
\bibitem{Tauc} J. Tauc, The share of thermal energy taken from the surroundings in the electro-luminescent energy radiated
from a p-n junction, Czech. J. Phys. {\bf 7}, 275 (1957).
\bibitem{Dousmanis} G. C. Dousmanis, C.W. Mueller, H. Nelson, and K. G. Petzinger, Evidence of refrigerating action by means of photon emission in semiconductor diodes, Phys. Rev. {\bf 133}, A316 (1964).
\bibitem{Berdahl} P. Berdahl, Radiant refrigeration by semiconductor diodes, J. Appl. Phys. {\bf 58}, 1369 (1985).
\bibitem{Santhanam} P. Santhanam, D. J. Gray Jr., and R. J. Ram, Thermoelectrically Pumped Light-Emitting Diodes Operating above
Unity Efficiency, Phys. Rev. Lett. {\bf 108}, 097403 (2012).
\end{thebibliography}
\end{document}